\documentclass[]{aa}
\usepackage{graphicx}
\usepackage{rotating}
\usepackage[varg]{txfonts}
\usepackage{natbib}
\usepackage[fleqn]{amsmath}
\usepackage[modulo, switch]{lineno}
\usepackage{nicefrac}
\usepackage{textcomp}
\usepackage{xfrac}
\usepackage{hyperref}

\usepackage[varg]{txfonts}
\def\mh{\,$\mu$Hz}
\def\teff{$T_{\mathrm{eff}}$}
\def\lg{\ensuremath{\log g}}

\def\num{$\nu_\mathrm{max}$}

\def\sun{\hbox{$_\odot$}}
\def\bc{BRITE\,-\,Constellation}

\begin{document}
%\linenumbers
%
   \title{Stellar masses from granulation and oscillations of 23 bright red giants observed by \bc \thanks{Based on data collected by the BRITE Constellation satellite mission, designed, built, launched, operated and supported by the Austrian Research Promotion Agency (FFG), the University of Vienna, the Technical University of Graz, the University of Innsbruck, the Canadian Space Agency (CSA), the University of Toronto Institute for Aerospace Studies (UTIAS), the Foundation for Polish Science \& Technology (FNiTP MNiSW), and National Science Centre (NCN).}}
% \subtitle{Stellar masses from the granulation and oscillation signal for 23 bright red giants}

\author{T. Kallinger\inst{1}
\and P.\,G. Beck\inst{2,3}
\and S. Hekker\inst{4,5}
\and D. Huber\inst{6}
\and R. Kuschnig\inst{7}
\and M. Rockenbauer\inst{1}
\and P.\,M. Winter\inst{1,8}
\and W.\,W. Weiss\inst{1}
\and G. Handler\inst{9}
\and A.\,F.\,J. Moffat\inst{10}
\and A. Pigulski\inst{11}
\and A. Popowicz\inst{12}
\and G.\,A. Wade\inst{13}
\and K. Zwintz\inst{14}
            }

   \offprints{thomas.kallinger@univie.ac.at}

   \institute{
Institut f\"ur Astrophysik, Universit\"at Wien, T\"urkenschanzstrasse 17, 1180 Vienna, Austria
        \and
Instituto de Astrof\'{\i}sica de Canarias, E-38200 La Laguna, Tenerife, Spain
	\and
Departamento de Astrof\'{\i}sica, Universidad de La Laguna, E-38206 La Laguna, Tenerife, Spain
	\and
Max Planck Institute for Solar System Research, Justus-von-Liebig-Weg 3, 37077 G\"ottingen, DE
	\and
Stellar Astrophysics Centre, Department of Physics and Astronomy, Aarhus University, Ny Munkegade 120, 8000 Aarhus C, DK
	\and
Institute for Astronomy, University of Hawai`i, 2680 Woodlawn Drive, Honolulu, HI 96822, USA
	\and
Institut f\"ur Kommunnikationsnetze und Satellitenkommunikation, Technical University Graz, Inffeldgasse 12, 8010 Graz, Austria
	\and
Institute for Machine Learning, Johannes Kepler University, Altenberger Str. 69, Computer Science Building, 4040 Linz, Austria
	\and
Nicolaus Copernicus Astronomical Center, ul. Bartycka 18, 00-716 Warsaw, Poland
	\and
D\'epartement de physique and Centre de Recherche en Astrophysique du Qu\'ebec (CRAQ), Universit\'e de Montr\'eal, CP 6128, Succ. Centre-Ville, Montr\'eal, Qu\'ebec, H3C 3J7, Canada
	\and
Instytut Astronomiczny, Uniwersytet Wroc{\l}awski, Kopernika 11, 51-622 Wroc{\l}aw, Poland
	\and
Institute of Automatic Control, Silesian University of Technology, Akademicka 16, 44-100 Gliwice, Poland
	\and
Department of Physics, Royal Military College of Canada, PO Box 17000 Station Forces, Kingston, ON, Canada K7K 0C6
	\and
Universit\"at Innsbruck, Institut f\"ur Astro- und Teilchenphysik, Technikerstrasse 25, A-6020 Innsbruck
}

   \date{Received  / Accepted}

\abstract
{The study of stellar structure and evolution depends crucially on accurate stellar parameters. The photometry from space telescopes has provided superb data that allowed asteroseismic characterisation of thousands of stars. However, typical targets of space telescopes are rather faint and complementary measurements are difficult to obtain. On the other hand, the brightest, otherwise well-studied stars, are lacking seismic characterization.
}
%Aims
{Our goal is to use the granulation and/or oscillation time scales measured from photometric time series of bright red giants (1.6\,$\leq$\,V\,mag\,$\leq$\,5.3) observed with \bc\ to determine stellar surface gravities and masses.}
%Methods
{We use probabilistic methods to characterize the granulation and/or oscillation signal in the power density spectra and the autocorrelation function of the \bc\ time series.}
%Results
{We detect a clear granulation and/or oscillation signal in 23 red giant stars and extract the corresponding time scales from the power density spectra as well as the autocorrelation function of the \bc\ time series. To account for the recently discovered non-linearity of the classical seismic scaling relations, we use parameters from a large sample of \textit{Kepler} stars to re-calibrate the scalings of the high- and low-frequency components of the granulation signal. We develop a method to identify which component is measured if only one granulation component is statistically significant in the data. We then use the new scalings to determine the surface gravity of our sample stars, finding them to be consistent with those determined from the autocorrelation signal of the time series. We further use radius estimates from the literature to determine the stellar masses of our sample stars from the measured surface gravities. We also define a statistical measure for the evolutionary stage of the stars.}
%Conclusions
{Our sample of stars covers low-mass stars on the lower giant branch to evolved massive supergiants and even though we can not verify our mass estimates with independent measurements from the literature they appear to be at least good enough to seperate high-mass from low-mass stars. Given the large known but usually not considered systematic uncertainties in the previous model-based mass estimates, we prefer our model-independent measurements.}

\authorrunning{Kallinger et al.}
\titlerunning{BRITE red giants}
   \maketitle

\section{Introduction}	\label{sec:intro}
During about the last 15 years, seismology of red giant stars has grown to an important field in stellar astrophysics, providing a unique opportunity to probe the interior structure of stars in an eventful phase of stellar evolution \citep[e.g., see reviews by][]{Chaplin2013,Hekker2017}.

This is possible thanks to so-called solar-type oscillations that are intrinsically damped and stochastically excited by the turbulent flux of the near-surface convection. For main-sequence stars, these are pressure (p) modes that are mainly confined to the outer envelope and which produce a comb-like pattern in the frequency spectrum. The dense He cores of red giants allow for the existence of gravity (g) modes at frequencies that overlap the p-mode frequency spectrum \citep[e.g.,][]{Dziembowski2001,Dupret2009}. Resonant coupling between the g- and p-mode cavities gives rise to so-called mixed modes, which carry valuable information about the deep interior of red giants but can be observed on their surface. 

Early attempts to observe solar-type oscillations in red giants used ground-based radial velocity measurements \citep[e.g.,][]{Frandsen2002,DeRidder2006} and space-based photometry (WIRE: \citealt{Buzasi2000,Retter2003,Stello2008}; HST: \citealt{Edmonds1996,kal2005,Stello2009,Gilliland2011}; SMEI: \citealt{Tarrant2007,Tarrant2008}; MOST: \citealt{Barban2007,kal2008b,kal2008a}). The major breakthrough, however, came with the launch of CoRoT \citep{Baglin2006} and \textit{Kepler} \citep{Borucki2010}, which provided data of unprecedented length and quality needed to tackle various questions in stellar astrophysics, like the relation between the different seismic \citep[e.g.][]{Huber2011,Mosser2012} and granulation \citep[e.g.][]{Mathur2011} observables, the analysis of non-radial mixed modes \citep[e.g.][]{Bedding2011,Beck2011,Mosser2012b} and how they are used to constrain the core-rotation rate \citep[e.g.][]{Beck2012,Mosser2012c,Gehan2018}, and the analysis of mode amplitudes and lifetimes \citep[e.g.][]{Baudin2011,vrard2018} and what they reveal about the signature of acoustic glitches \citep[e.g.][]{Corsaro2015,Vrard2015}. 

Moreover, seismology of red giants resulted in a much more precise method for determining stellar global parameters compared to classical methods such as isochrone fitting \citep[e.g.][]{Lebreton2014}. The determination of accurate stellar parameters is a fundamental and longstanding problem in astrophysics \citep[e.g.][]{Soderblom2010}. However, this is often only possible by means of stellar models and therefore suffers from our incomplete knowledge of the physical processes inside stars. The wealth of observations delivered by CoRoT and \textit{Kepler} triggered the development of a new approach, which is founded on relations between global seismic parameters, such as the large frequency separation or the frequency of the maximum oscillation power, and stellar parameters \citep{Ulrich1986,Brown1991,Kjeldsen1995}. Initially, these scaling relations were used to predict the characteristics of the oscillations. \cite{Stello2008} and \cite{kal2010a} were the first to obtain seismic masses and seismic masses and radii, respectively from global oscillation parameters. Mainly empirical, the seismic scaling relations have been shown to be accurate throughout many evolutionary stages \citep[e.g.][]{White2011} and are nowadays widely used in various fields, from determining fundamental parameters of field stars \citep[e.g.][]{kal2010,Mosser2010}, binary systems \citep[e.g.][]{Beck2014,Gaulme2016} and planet-hosting stars \citep[e.g.][]{Borucki2012} to Galactic population studies \citep{Miglio2013}.

However, the calibration of the seismic scalings still remains to be an issue. Several studies have shown that the extrapolation from the Sun to evolved red giants causes biases and yields too high masses for red giants \citep[e.g.][]{Miglio2012,Gaulme2016}. Various model-based corrections \citep[e.g.][]{White2011a,Sharma2016} yield better results but suffer from known deficiencies in the models. \cite{kal2018} defined revised scaling relations that account for these systematic discrepancies in a completely model-independent way.  

\begin{figure}
	\begin{center}
	\includegraphics[width=0.5\textwidth]{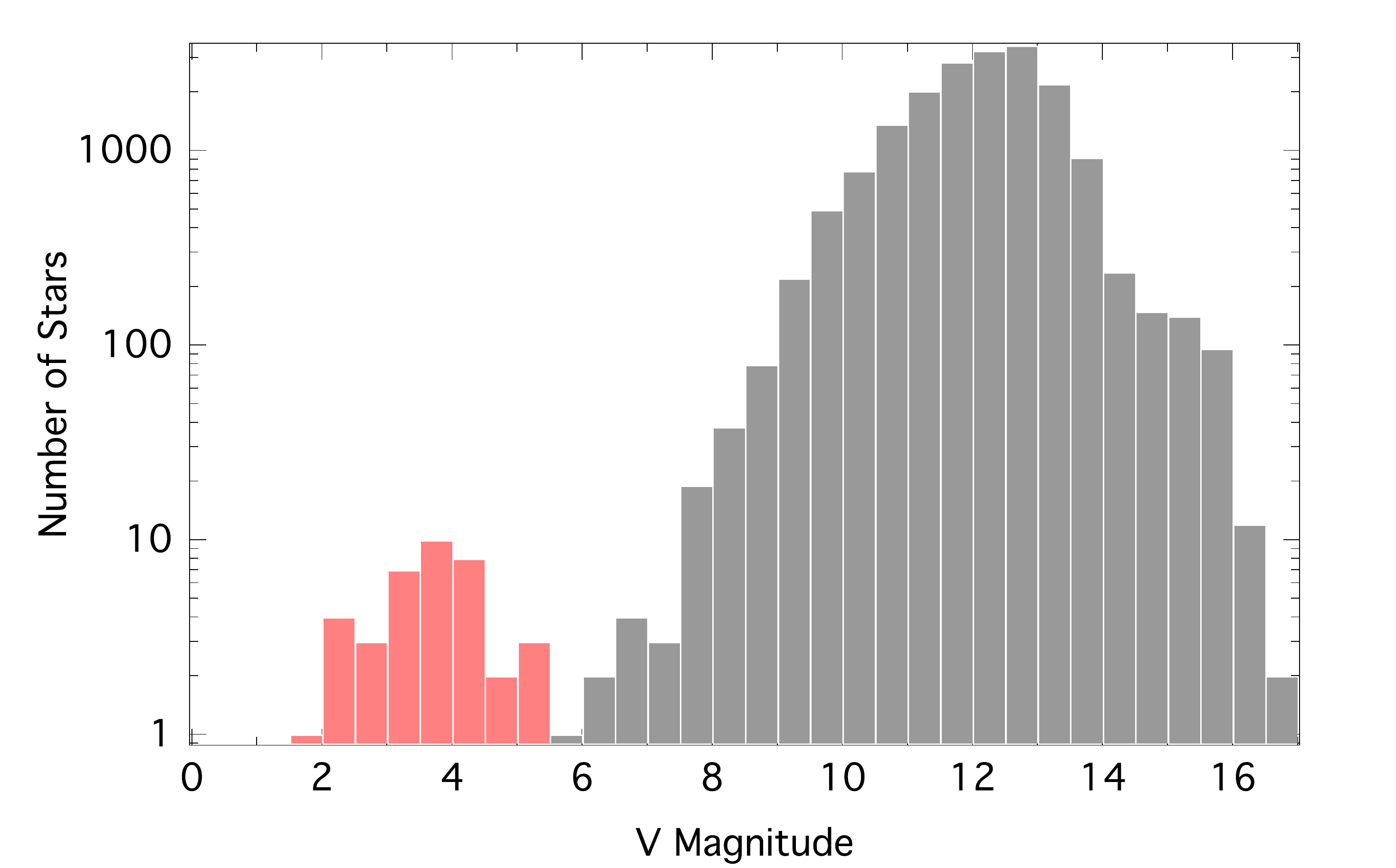}
	\caption{Histrogram of magnitudes of red giant stars observed by \bc\ (red) and \textit{Kepler} \citep[grey;][]{Huber2014}.} 
	\label{fig:histo} 
	\end{center} 
\end{figure}

From an observational point of view, the CoRoT and \textit{Kepler} missions have contributed tremendously to the field of asteroseismology of red giants and even though they provided data of unprecedented quality they also suffer from the fundamental problem that they are essentially limited to relatively faint stars (see Fig.\,\ref{fig:histo}). For such stars it is difficult to get additional constraints from, e.g. high-resolution spectroscopy, interferometry, etc. \citep[e.g.][]{Huber2012}. For bright red giants, on the other hand, there are apart from a few exceptions \citep[see earlier references and e.g.][]{Pope2016} basically no photometric time-series data available that allow for detailed asteroseismic analyses. The TESS mission \citep{Ricker2015} will change this but even though already operating it still will need several months to accumulate the necessary observing lengths. In the meantime, the satellites of the BRITE (BRIght Target Explorer) - Constellation\footnote{\href{url}{http://www.brite-constellation.at}} mission can efficiently help to fill this gap. The BRITE satellites are designed to photometrically monitor the brightest stars in the sky for up to six months and therefore provide the time base needed for useful asteroseismic analyses of red giants. Even though the photometric quality of the BRITE instruments is likely not sufficient to clearly detect solar-type oscillations on the lower red-giant branch (with typical amplitudes of some 10 to some 100\,ppm), a large fraction of the potential BRITE targets (i.e., stars brighter than about 5th magnitude) are red giants. Therefore, it was decided to monitor these stars whenever they are present in an observing field without disturbing other ongoing programs. 

In this paper we report on the \bc\ observations of red giants during the  first three years of the mission. During this time the BRITE satellites collected photometric time series data for 38 red giants consisting of more than 2 million individual measurements. We find a granulation and/or oscillation signal in 23 of the 38 stars located in 9 of the 15 fields observed. This sample covers a large variety of red giants ranging from low-mass red clump stars to high-mass red supergiants. We extract the typical granulation and/or oscillation timescales of these stars from which we determine accurate surface gravities. Using radius estimates from the literature (mainly based on interferometric angular diameters and Gaia parallaxes) we can constrain the masses of these stars with typical uncertainties of 10-15\%. We obtain consistent results using different methods for our measurements and mass estimations.

%--------------------------------------------------------------------
%--------------------------------------------------------------------

\section{BRITE photometry}\label{sec:phot}
The photometric observations used in this study were collected with \bc . This is a fleet of five nanosatellites described in detail by \cite{Weiss2014} and \cite{pablo2016} and consists of two Austrian satellites UniBRITE (UBr) and BRITE-Austria (BAb), one Canadian satellite BRITE-Toronto (BTr), and two Polish satellites BRITE-Heweliusz (BHr) and BRITE-Lem (BLb). Each of the 20$\times$20$\times$20\,cm satellites hosts an optical telescope of 3\,cm aperture, feeding an uncooled CCD, and is equipped with a single filter. Three nanosats (UBr, BTr, BHr) have a red filter (550--700\,nm) and two (BAb, BLb) have a blue filter (390--460nm), which are similar to the Sloan $r$ and Geneva $B$ filters, respectively. 

The observing strategy of \bc\ is to point at least one satellite at a $20\degr \times 24\degr$ field of the sky and simultaneously collect data for about 30 stars in this field, which are visible for about 5--30\,min per 97--101\,min orbit for up to six months.  

A problem affecting the BRITE satellites is the sensitivity of the Kodak KAI-11002M detectors (11 million pixels with a plate scale of 27.3\arcsec\ per pixel) to particle radiation, which posed a major threat to the lifetime and effectiveness of the BRITE mission. The impact of high-energy protons causes the emergence of hot and warm pixels at a rate much higher than originally expected. The affected pixels more easily generate thermal electrons and thereby significantly impair the photometric precision of the observations. An additional important problem that appeared after several months of operation was the charge transfer inefficiency, also caused by the protons. Thanks to increasing the readout time and adopting a \textit{chopping} technique for data acquisition, the effect of CCD radiation damage on the photometry is now significantly reduced \citep{Popowicz2017}. 

In the chopping mode, the satellite pointing is shifted slightly between consecutive exposures, so that the target point spread functions (PSF) centroids alternate by about 20 pixels between two positions on the CCD. This means that the PSF-free part of a given subraster image acts as a dark image for the subsequent exposure and subtracting consecutive exposures results in an image with one negative and one positive target PSF. The background defects are thereby almost entirely removed.

%________________________________________________________________Tab. BRITE Fields
\begin{table}[t]
\begin{small}
\centering
\caption{Summary of the observed fields. The columns \#, $d$, and M list the number of observed stars that were observed for $d$ days with observing mode S (stare) or Ch (chopping), respectively.
\label{tab:fields}}
\begin{tabular}{lcccccc}
\hline\hline
\noalign{\smallskip}
Field & Sat. &Start date&End date&d&\#&M\\
\noalign{\smallskip}
\hline
\noalign{\smallskip}
\multicolumn{7}{l}{Cyg I}\\
	&UBr	&2014-06-12	&2014-07-01	&19	&24	&S\\	
	&BLb	&2014-07-12	&2014-11-24	&135&22	&S\\	
	&BTr	&2014-07-06	&2014-09-23	&80	&31	&S\\	
\multicolumn{7}{l}{Per I}\\
	&UBr	&2014-09-04	&2015-02-18	&168	&33	&S/Ch\\	
	&BAb	&2014-09-13	&2015-01-23	&132	&19	&S\\	
\multicolumn{7}{l}{Vel$-$Pup I}\\
	&BAb	&2014-12-10	&2015-05-23	&164	&31	&S/Ch\\
	&BTr	&2014-12-19	&2015-05-27	&159	&36	&S/Ch\\
\multicolumn{7}{l}{Sco I}\\
	&UBr	&2015-03-20	&2015-08-29	&162	&18	&Ch\\
	&BAb	&2015-03-28	&2015-07-19	&113	&8	&Ch\\	
	&BLb	&2015-03-17	&2015-08-26	&162	&19	&Ch\\	
	&BHr	&2015-06-26	&2015-08-28	&64		&18	&S/Ch\\	
\multicolumn{7}{l}{CMa$-$Pup I}\\
	&BLb	&2015-10-26	&2016-04-10	&167	&26	&Ch\\
	&BTr	&2015-11-15	&2016-04-18	&157	&21	&Ch\\
	&BHr	&2015-10-27	&2016-04-14	&169	&17	&Ch\\
\multicolumn{7}{l}{Cru$-$Car I}\\
	&UBr	&2016-03-06	&2016-07-22	&139	&21	&Ch\\	
	&BAb	&2016-02-04	&2016-05-27	&113	&19	&Ch\\	
	&BLb	&2016-02-19	&2016-07-15	&147	&19	&Ch\\	
	&BTr	&2016-02-09	&2016-07-22	&165	&27	&Ch\\	
	&BHr	&2016-02-19	&2016-04-27	&68		&18	&Ch\\	
\multicolumn{7}{l}{Sgr II}\\
	&BAb	&2016-04-21	&2016-09-13	&145	&12	&Ch\\
	&BHr	&2016-04-28	&2016-09-20	&146	&19	&Ch\\
\multicolumn{7}{l}{Cet$-$Eri I}\\
	&BHr	&2016-10-06	&2017-01-04	&91		&28	&Ch\\
\multicolumn{7}{l}{Car I}\\
	&UBr	&2017-01-11	&2017-05-04	&113	&21	&Ch\\
	&BAb	&2017-01-11	&2017-02-13	&33	&19	&Ch\\
	&BHr	&2017-01-29	&2017-07-01	&154	&27	&Ch\\
\hline
\end{tabular}
\end{small}
\end{table}
%__________________________________________________________________

Details about data reduction are given by \cite{Popowicz2017}. While in stare observing mode, bias, dark, and background corrections are necessary, they are not required for observations in chopping mode since the stellar flux is extracted from differential images. The remaining main step is to identify the optimal apertures and to extract the flux within these apertures. The light curves resulting from this pipeline reduction are deposited in the \bc\ data archive from where we extract the data.

\section{The target sample}
Between 2013 and 2017, \bc\ has observed 24 fields. In 15 of them, a total of 38 red giant stars were observed (see Fig.\,\ref{fig:histo}). We carefully examined these data and find significant intrinsic variability in 23 stars observed in nine different fields. Two stars were observed in two campaigns. The sample includes a large variety of red giants with apparent magnitudes between 1.6 and 5.3\,mag and spectral types ranging from G2\,III to M3.5\,II. For the remaining 15 stars the data sets are either too short or too noisy to detect variability.

Several stars were observed by more than one satellite. However, in this work we only show results from the data set with the best quality. A summary of the different fields observed is given in Tab.\,\ref{tab:fields}, while details about the individual stars are listed in Tab.\,\ref{tab:stars}.

%________________________________________________________________Tab. BRITE Red Giants
\begin{sidewaystable*}
%\begin{table}[t]
\begin{small}
\centering
\caption{Overview of the 23 target stars. Details about the observed fields are given in Tab.\,\ref{tab:fields}. Effective temperatures and luminosities are taken from \cite{McDonald2017}.  The observational durations are listed for all satellites that observed a given star, with the data set used in this study marked in bold-face.  The column ``data points'' refers to the number of data points in the reduced and subsequently binned final data set. The column ``duty cycle'' gives the duty cycle of the binned and gap-filled time series. Stars marked with $^\dag$ are flagged as variable stars in the General Catalog of Variable Stars \citep{GCVS2017}. For $\rho$\,Per and $\varepsilon$\,Mus the catalog lists a period of 50 and 40\,d, respectively.
\label{tab:stars}}
\begin{tabular}{l|ll|cccccc|c|rrrrr|cccc}
\hline\hline
\noalign{\smallskip}
ID&\multicolumn{2}{c|}{Name} &   V   &   Sp.\,Type    &   $T_\mathrm{eff}$  &   $\log g$ & $\log L$	&Fe/H  &Field &  Ubr   & BAb &     BHr&   BLb &  BTr   & \multicolumn{2}{c}{data points}& \multicolumn{2}{c}{duty cycle [\%]}	\\
&&& [mag]& &  [K] & [cm/s$^2$] &[L\sun]&&& \multicolumn{5}{c}{[d]} & red. & bin. &  bin. & gap-filled\\
\hline
\noalign{\smallskip}
A&   $\eta$\,Cyg   & \object{HD 188947}  		&    3.89   &   K0 III      &   4872$\pm$125   &   2.6$^b$ & 1.78$\pm$0.01$^b$ &$-0.02^c$&Cyg I &   19   &            &     &32 &   \bf{80}  & 25069 &759 & 65 & 76	\\
B&  23\,Vul    & \object{HD 192806}  			&    4.52   &   K3 III      &   4333$\pm$125   &   2.0$^d$ & 2.46$\pm$0.02$^b$ &$-0.22^d$&Cyg I &        &      &          & &      \bf{74}  &  22412 &	687 & 64 & 74\\
C&  39\,Cyg    & \object{HD 194317}  			&    4.43   &   K2.5 III    &   4284$\pm$125   &   1.8$^e$ & 2.27$\pm$0.01$^b$ &$0.04^e$&Cyg I &  19    &            &     & &     \bf{80}  & 38555 &770 & 66 & 76	\\
D&  $\varepsilon$\,Cyg    & \object{HD 197989}  	&    2.46   &   K0 III    &   4740$\pm$125   &   2.6$^c$ & 1.75$\pm$0.02$^a$&$-0.10^c$ &Cyg I &  20    &            &     & 126&     \bf{48}  & 15408 &481 & 67 & 79	\\

E& $\rho$\,Per$^\dag$   & \object{HD 19058}   			&    3.39   &   M4 II      &   3479$\pm$125   &   0.8$^d$ &3.43$\pm$0.03$^a$ &$-0.15^d$ &Per I &   \bf{162}  &        & &           &    & 136091 &1933 & 80  & 94 	\\
F& $\sigma$\,Per   & \object{HD 21552}   		&    4.36   &   K3 III      &   4206$\pm$125   &   1.9$^d$ &2.54$\pm$0.01$^b$ &$-0.24^d$ &Per I &   \bf{162}  &     & &              &    & 134772 &1909 & 79  & 85 	\\

G&  NS\,Pup    & \object{HD 68553}   		&    4.45   &   K3 Ib      &   3961$\pm$125   &   0.2$^a$ &3.99$\pm$0.08$^a$ &-&Vel$-$Pup I &        &            &    & &      \bf{69}   & 17053 &756 & 75 & 88	\\
H&  $\beta$\,Pyx  & \object{HD 74006} 	&    3.97   &   G7 Ib-II      &   5124$\pm$125   &   1.6$^a$ &2.51$\pm$0.02$^b$ &0.25$^d$&Vel$-$Pup I &             &   2    & &      &   \bf{69}  & 17155 &749 & 74 & 88	\\
I&  g\,Car     & \object{HD 80230}   		&    4.34   &   M1 III      &   3834$\pm$125   &   0.7$^a$ &3.07$\pm$0.02$^b$ &-&Vel$-$Pup I &        &      &         & &       \bf{61}  & 20849 &643 & 64 & 75	\\
      &    &       &        &      &     & && &Car I  &   \bf{76}   & &     &       &     &  41698 &623 & 57  & 76	\\
J&  N\,Vel$^\dag$     & \object{HD 82668}   		&    3.13   &   K5 III      &   3964$\pm$125   &   1.0$^a$ &2.89$\pm$0.04$^b$ &-&Vel$-$Pup I &        &    &        & &        \bf{69}   & 17019 &763 & 76	 & 88\\
      &    &       &         &      &    &  &&&Car I &  \bf{76}        & &        &       &      & 21189 &624 & 57 & 75	\\
K&  $\varepsilon$\,Sco     & \object{HD 151680}   	&    2.29   &   K2.5 III      &   4583$\pm$125   &   2.3$^a$ &1.73$\pm$0.02$^a$ &$-0.17^d$&Sco I &   \bf{161}     &    &        & &          & 57712 &1201 & 53	 & 81\\

L& $\sigma$\,CMa$^\dag$   & \object{HD 52877}	&   3.47   &   K7 Ib       &   3792$\pm$125   &   1.0$^d$ &4.09$\pm$0.04$^a$ &0.16$^d$&CMa$-$Pup I&        &    &    \bf{173}  &      &     & 58838 & 1695 & 68   & 74	\\

M&  $\varepsilon$\,Mus$^\dag$   & \object{HD 106849} &    4.11   &   M5 III      &   3470$\pm$125   &   0.3$^a$ &3.24$\pm$0.03$^a$ &-&Cru$-$Car I  &   \bf{134}     & &   68    &       &       &35231 &1395 & 71 & 76	\\
N&  $\varepsilon$\,Cru   & \object{HD 107446}  &    3.59   &   K3 III      &   4210$\pm$125   &   1.8$^d$ &2.45$\pm$0.01$^b$ &$-0.03^d$&Cru$-$Car I  &   \bf{134}    & &   68     &       &       &41895 &1384 & 70 & 76	\\
O&  $\gamma$\,Cru   & \object{HD 108903} &    1.63   &   M3.5 III    &   3689$\pm$125   &   0.7$^a$ &2.92$\pm$0.03$^a$ &-&Cru$-$Car I  &   \bf{134}    & 139& 68    & 139   &      & 42185 &1400 & 71 & 76 	\\

P&  G\,Sco     & \object{HD 161892}  			&    3.21   &   K2 III      &   4535$\pm$125   &   2.2$^f$ &1.97$\pm$0.02$^a$  &-&Sgr II &         &     &  \bf{142}    &    &        & 85459 &1461 & 67 & 74	\\
Q&  $\gamma^2$\,Sgr  & \object{HD 165135}  	&    2.99   &   K0 III      &   4834$\pm$125   &   2.7$^d$ &1.78$\pm$0.02$^a$ &$-0.36^d$ &Sgr II &      &        &  \bf{142}   &     &      &85683 & 1460 & 67 & 74 	\\
R&  $\eta$\,Sgr$^\dag$   & \object{HD 167618}  		&    3.11   &   M3.5 III    &   3638$\pm$125   &   0.9$^a$ &2.72$\pm$0.03$^a$&-  &Sgr II &        &      &  \bf{142}  &      &   &   85483 &1451 & 67 & 74 	\\

S&            & \object{HD 19349}   			&    5.26   &   M2.5 II    &   3556$\pm$125   &   0.7$^a$ & 3.11$\pm$0.03$^b$ &-&Cet$-$Eri I &        &    &     \bf{90}    &    &      & 31661 & 631 & 48 & 61 	\\
T&  $\alpha$\,Cet   & \object{HD 18884}   	&    2.53   &   M2 III      &   3890$\pm$150   &   1.3$^d$ & 3.19$\pm$0.05$^a$ &0.00$^c$&Cet$-$Eri I &        &     &    \bf{90}    &    &     & 29565 &642 & 49 & 60  	\\

U&  V337\,Car$^\dag$  & \object{HD 89388}   			&    3.35   &   K3 IIa      &   4118$\pm$125   &   1.6$^d$ & 3.51$\pm$0.03$^a$ &0.54$^d$&Car I  &   \bf{102}  &       & &            &   & 27241 &842 & 56 & 71    	\\
V&  GZ\,Vel$^\dag$    & \object{HD 89682}   			&    4.54   &   K3 II      &   4079$\pm$125   &   1.5$^g$ &3.73$\pm$0.04$^b$  &-&Car I  &   \bf{76}   &      &      & &            & 21559& 621 & 57  & 75 	\\
W&  12\,Mus    & \object{HD 101379} 			 &    5.01   &   G2 III      &   4744$\pm$125   &   1.7$^a$ & 2.10$\pm$0.04$^b$  &-&Car I  &       &       & \bf{152}     &    &      & 81444& 1700 & 74 & - 	\\
\hline
\end{tabular}
\tablefoot{
Values for $T_\mathrm{eff}$ are taken from \cite{McDonald2017}. Values for $\log g$, $\log L$, and Fe/H are from:
\tablefoottext{a}{\cite{McDonald2017}}--
\tablefoottext{b}{\cite{GaiaDR2}}--
\tablefoottext{c}{\cite{Prugniel2007}}--
\tablefoottext{d}{\cite{Soubiran2016}}--
\tablefoottext{e}{\cite{Reffert2015}}--
\tablefoottext{f}{\cite{Allende1999}}--
\tablefoottext{g}{\cite{Borde2002}}\\
}
\end{small}
%\end{table}
\end{sidewaystable*}
%__________________________________________________________________

\cite{McDonald2017} derived consistent estimates of the effective temperatures for all stars in our sample based on Hipparcos-Gaia parallaxes and a comparison between archival multi-wavelength photometry and stellar model atmospheres. Stellar luminosities are taken from the Gaia DR2 catalogue \citep{GaiaDR2} and from \cite{McDonald2017}. For our target stars the uncertainties are typically $\pm$125\,K for the effective temperature and 5--10\% for the luminosity. Literature values for the surface gravity and metallicity are less consistent and are taken from various sources with typically much larger uncertainties. 

\section{BRITE data post-processing}\label{sec:postproc}

The photometric time series as delivered by the data reduction team still includes instrumental effects and obvious outliers and therefore needs some post-processing. The \bc\ data come in different setups\footnote{``Setup'' refers to a set of camera parameters but also to subraster positions on the CCD. Setups may change at the beginning of a run (during optimising of the observations) or due to adding/removing a star during the run. For each parameter change a whole new setup is generated with a particular ID.} and data blocks of approximately equal length, which we treat independently. Subdivision of the data sets into blocks was required due to a limit of typically 30\,000 frames for the standard data reduction software.

%--------------------------------------------------------------------
\begin{figure}
	\begin{center}
	\includegraphics[width=0.5\textwidth]{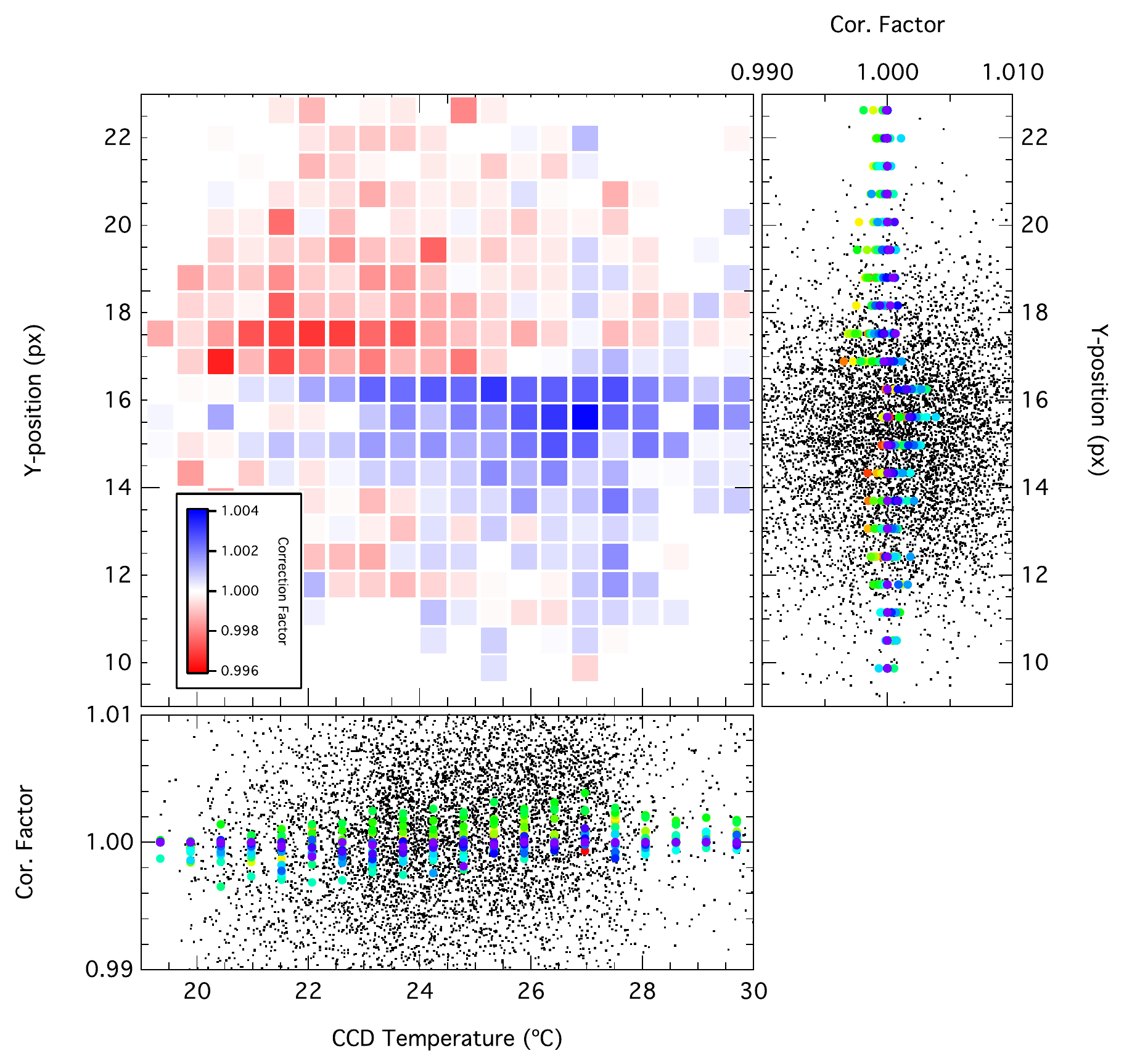}
	\caption{2D correction map (CCD-Temperature and Y position of the PSF) for the about 30 day-long UBr time series of $\varepsilon$ Sco obtained during observing subset 2 with the colour-coded correction factors. The right and bottom panels show the correction factor as a function of the Y position of the PSF and the CCD temperature, respectively, where the values of the other parameter are colour-coded (with red-green-blue symbols corresponding to low-mean-high values of the other parameter). The black dots are the actual measurements divided by the average value of the time series.} 
	\label{fig:cormap} 
	\end{center} 
\end{figure}
%--------------------------------------------------------------------

Adapting the recipe of \cite{pigulski2016} and \cite{kal2017} we perform the following steps for each block of each data set: 
\begin{itemize}
  \item If the data were obtained in chopping mode we divide the data set into two sets corresponding to the alternating PSF position on the CCD subraster. 
  \item Compute a 2D histogram of the X/Y positions and fit a 3D multivariate Gaussian to it. Images that were taken with the PSF centre positioned outside three times the widths of the Gaussian are eliminated from further processing. This procedure identifies most of the outliers and rejects them.
 \item The procedure continues with the ``cleaned" data set and applies a 4$\sigma$-clipping. Depending on the amplitude of the intrinsic signal the $\sigma$-clipping is performed relative to the average flux of the complete data set or a running average with adaptable width.
 \item Compute the correlation coefficient $c$ between the instrumental flux and the CCD temperature and X/Y position of the PSF.
 \item Build an $n$-dimensional correction map for all parameters with $|c|>0.1$ (i.e., a 1D, 2D, or 3D map), where each axis is divided into $N^{1/(n+1)}$ bins, where $N$ is the number of data points, which results in typically 10-30 bins. All measurements in a given bin are then averaged. Finally, the correction map is divided by the average instrumental flux to receive a correction factor for each combination of instrumental parameters, which is then applied to the time series. As an example, we show in Fig.\,\ref{fig:cormap} the resulting 2D correction map (CCD temperature and Y position of the PSF) of a subset of the UBr observations of $\varepsilon$ Sco, which is the star with the lowest Fourier noise in our sample. The time series has an average instrumental flux of about $1.1\times 10^5$\,ADU/s so that a correction factor of 0.4\% corresponds to an instrumental signal of 440ADU/s. The rms scatter of the time series reduces from $\sim$6.4\,ppt before the correction to $\sim$5.4\,ppt after the correction, improving the photometric quality by about 16\%.
\end{itemize}

Finally, the data subsets are combined and divided by the average value for conversion to relative flux. The 23 final light curves are between 48 and 173\,d long and consist of about 15\,000 to 136\,000 individual measurements (see Tab.\,\ref{tab:stars}). 

\begin{figure*}
	\begin{center}
	\includegraphics[width=1.0\textwidth]{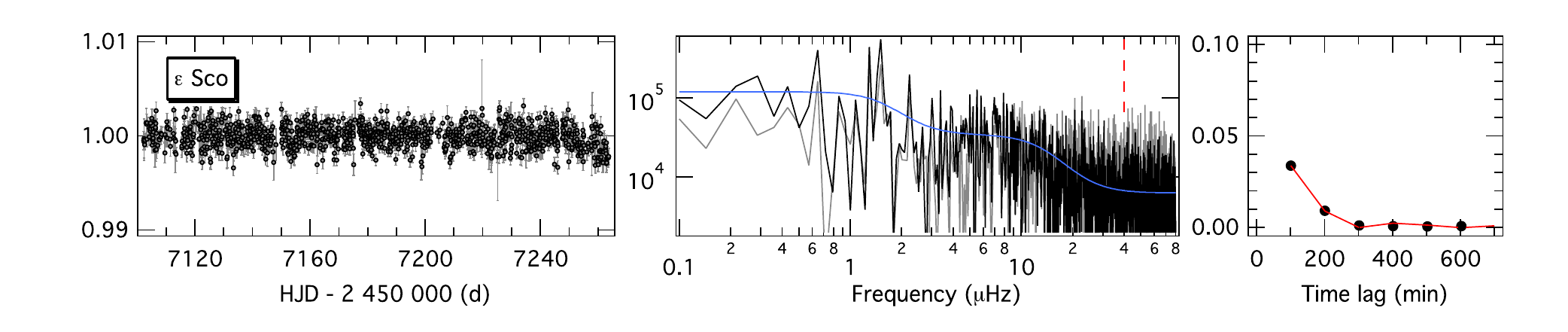}
	\caption{The left panel shows the relative instrumental flux of the about 161 day-long UBr binned data set of $\varepsilon$\,Sco. The middle panel shows the power density spectrum (in ppm$^2/\mu$Hz) of the binned (grey lines) and interpolated time series (black lines) along with a best global model fit (blue lines) and expected \num (vertical red lines). The right panel indicates the squared autocorrelation function of the binned time series (black dots) and a best fit (red line).} 
	\label{fig:stars0} 
	\end{center} 
\end{figure*}

\begin{figure*}
	\begin{center}
	\includegraphics[width=1.0\textwidth]{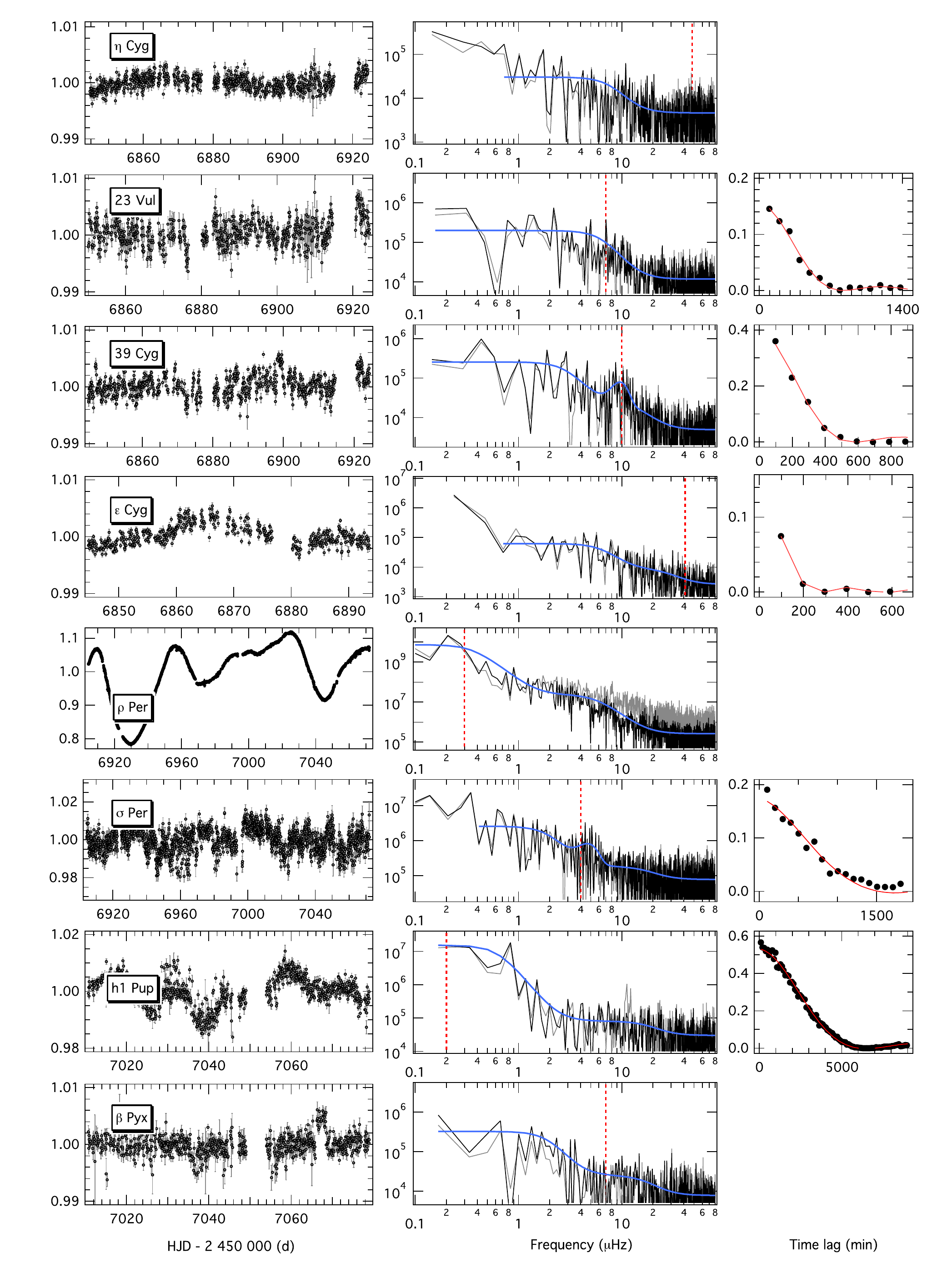}
	\caption{Same as Fig.\,\ref{fig:stars0} -- in case of $\eta$\,Cyg, $\rho$\,Per, and $\beta$\,Pyx the ACF signal is not conclusive.} 
	\label{fig:stars1} 
	\end{center} 
\end{figure*}

\begin{figure*}
	\begin{center}
	\includegraphics[width=1.0\textwidth]{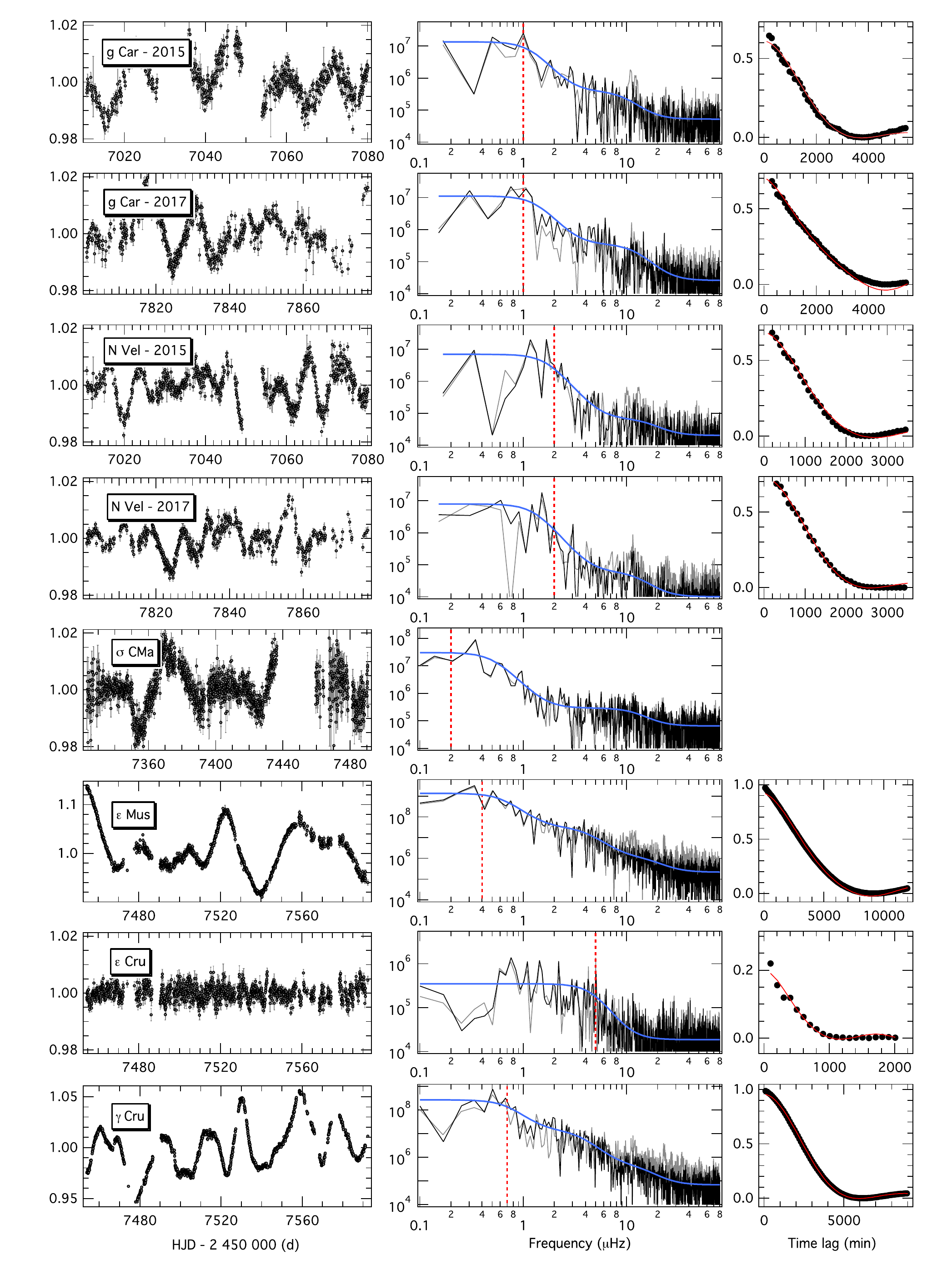}
	\caption{Same as Fig.\,\ref{fig:stars0}} 
	\label{fig:stars2} 
	\end{center} 
\end{figure*}

\begin{figure*}
	\begin{center}
	\includegraphics[width=1.0\textwidth]{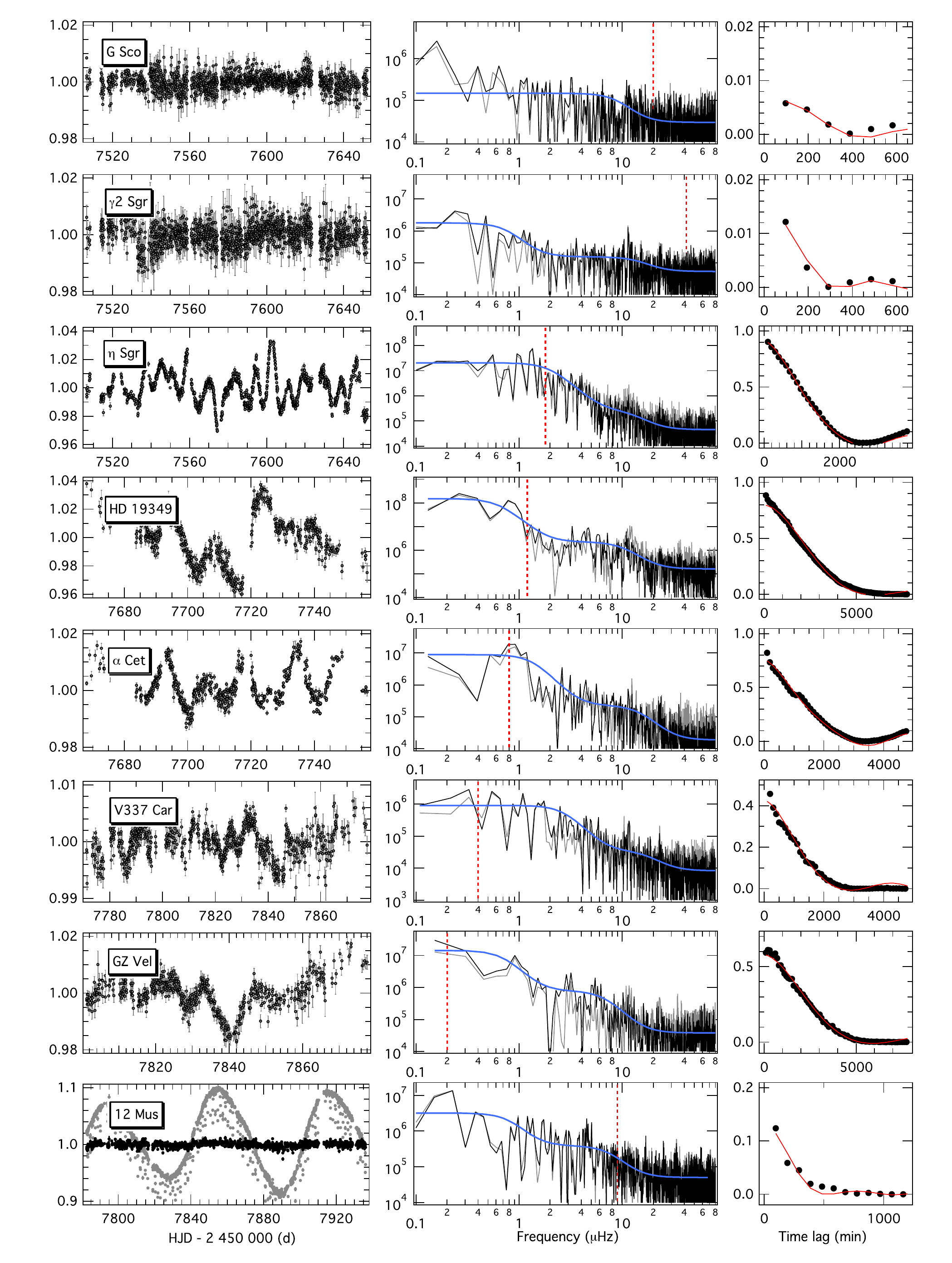}
	\caption{Same as Fig.\,\ref{fig:stars0}. For 12\,Mus we correct the original light curve for the long-period variation and eclipses originating from two companions.} 
	\label{fig:stars3} 
	\end{center} 
\end{figure*}

From the known fundamental parameters of our target stars we can expect that the intrinsic variability of all stars acts on time scales not shorter than a few hours. Hence, the typical cadences of the observations of some twenty seconds are unnecessarily short. We therefore bin the data into one measurement per BRITE orbit (97--101\,min, depending on the satellite), where the standard deviation of the original measurements within a given bin ($\sigma_\mathrm{bin}$) provides a good estimate for the photometric accuracy. The resulting Nyquist frequencies range from $\nu_\mathrm{Nyq}=82.5$ to 85.9\mh , which is in any case high enough to cover the full expected intrinsic variability of our target stars. The binned light curves consist of 481 to 1933 data points. The corresponding duty cycles (i.e., the actual number of data points relative to the total number of satellite orbits during the full length of the time series) range between 48 and 80\%. As an example, we show the fully reduced and binned light curve of $\varepsilon$\,Sco in Fig.\,\ref{fig:stars0}. The light curves of the other stars are illustrated in Fig.\,\ref{fig:stars1}--\ref{fig:stars3}.

In a next step we compute the unweighted Fourier spectrum of the time series and convert it to power density by dividing the spectral power by a factor that results from Parseval's theorem. The power density spectrum (PDS) of $\varepsilon$\,Sco is shown in Fig.\,\ref{fig:stars0} and that of the other target stars in Fig.\,\ref{fig:stars1}--\ref{fig:stars3}. 

%--------------------------------------------------------------------
%--------------------------------------------------------------------

A challenge for the subsequent analysis is the relatively low duty cycle of the \bc\ observations. For stars with a strong long-periodic signal the spectral window leaks power from the low-frequency into the high-frequency domain. If the contrast between intrinsic low-frequency and high-frequency signal is high enough, the high-frequency domain can then be dominated by this leaked power, which would significantly influence a global fit to the spectrum (see Sec.\,\ref{sec:data_analysis}). To improve the situation we fill gaps that are shorter than 0.5\,d by linear interpolation at a cadence that corresponds to the BRITE orbital period (i.e., up to seven data points for a 0.5\,d-long gap). Such an approach is well-tested for \textit{Kepler} observations \citep[e.g.,][]{Garcia2014,kal2014,Stello2017} and only marginally affects the frequency range we are interested in but significantly improves the duty cycle (see Tab.\,\ref{tab:stars}) and therefore the spectral window function. We then compute the power density spectra of the gap-filled time series. As can be seen in Fig.\,\ref{fig:stars0}, the power density of $\varepsilon$\,Sco significantly decreases above about 20\mh , while the low-frequency domain remains basically the same. The most severe case in our sample is $\rho$\,Per (Fig.\,\ref{fig:stars1}) for which the ``noise'' at high frequencies decreases by about an order of magnitude due to gap-filling in the time series (improving the duty cycle from 80 to 94\%). Another effect that we find is that for several stars (e.g., NS\,Pup) peak amplitudes around 1 and/or 2\,d$^{-1}$ ($\sim$11.6 and 23.1\mh ) are damped by gap-filling.

\section{Data analysis}	\label{sec:data_analysis}
Brightness variations in stars with a convective envelope arise primarily from granulation and acoustic oscillations \citep[e.g.,][]{Mathur2011,kal2014}, which are largely gravity-dependent. Further processes contributing to the photometric variations are dark and bright spots due to magnetic activity, which are modulated by the period of stellar rotation \citep[e.g.,][]{Garcia2014b,Ceillier2017,Beck2018}. Rotation and magnetic activity are largely gravity-independent. While these phenomena act on different time and amplitude scales for low-luminosity red giants, it might become more difficult to disentangle them with growing luminosity. However, for an a priori unknown star it is difficult to assign which signal is due to which phenomenon.

%--------------------------------------------------------------------
%--------------------------------------------------------------------
%--------------------------------------------------------------------
%--------------------------------------------------------------------
%--------------------------------------------------------------------
%--------------------------------------------------------------------

\subsection{Granulation or oscillation time scales}

\cite{kal2014} have shown that the power density spectrum (PDS) of a red giant (as for any other star with a convective envelope) can be globally modelled by a sequence of super-Lorentzian functions and a Gaussian. They found that the granulation signal is split into two components with their characteristic frequencies strongly tied to the frequency of maximum oscillation power (\num , which is usually defined as the centre of the Gaussian) and approximately scale as,
\begin{equation}		\label{eq:nugran}
\nu_\mathrm{low}=0.32\,\nu_\mathrm{max}^{0.97}	\,\,\,\,\,\,\,\,\,\,\,\,\,\,\,\,	\nu_\mathrm{high}=0.95\,\nu_\mathrm{max}^{0.99},
\end{equation}
where $\nu_\mathrm{low}$ and $\nu_\mathrm{high}$ are the characteristic frequency of the low-frequency (LF) and the high-frequency (HF) component, respectively.

The surface gravity $g$ can be estimated from a measure of $\nu_\mathrm{low}$, $\nu_\mathrm{high}$, or \num , if the effective temperature of a star is known, according to: 
\begin{equation}	\label{eq:numax0}
\nu_\mathrm{max} \propto g/\sqrt{T_\mathrm{eff}}
\end{equation}
\citep[e.g.,][]{Brown1991,Kjeldsen1995}.

In order to determine one (or more) of these parameters we follow the approach of \cite{kal2014} and fit the observed power density spectra with a global model:
\begin{equation}	\label{eq:globalmodel}
P(\nu) = P_n + \eta(\nu)^2 \left [ \sum_i \frac{2\sqrt{2}/\pi a_{i}^2/\nu_{i}}{1+(v/\nu_{i})^4} + P_g \exp \frac{-(\nu - \nu_\mathrm{max})^2}{2\sigma^2} \right ],
\end{equation}
where $P_n$ corresponds to the instrumental noise contribution and $a_i$ and $\nu_i$ to the rms amplitude and characteristic frequency of the $i$th component, respectively. $P_g$, \num , and $\sigma$ are the height, central frequency, and width of the oscillation power excess, respectively. The function $\eta(\nu)$ accounts for the frequency-dependent damping of the signal amplitude due to averaging over the integration time \citep{Huber2010} and is defined as,
\begin{equation}
\eta(\nu) = \mathrm{sinc} \left ( \frac{\pi \nu}{2\nu_\mathrm{Nyq}} \right ).
\end{equation}
For the fit we use the Bayesian nested sampling algorithm \textsc{MultiNest} \citep{Feroz2009}, which provides the posterior probability distributions for the parameter estimation and the global model evidence. The big advantage of the latter compared to other statistical tools is that it is properly normalised and evaluated over the entire parameter space and therefore allows one to reliably rate how well a given model represents a given data set compared to another model with little risk to over-fit the data. More details are given by \cite{kal2012,kal2014}.

Most of the stars in our sample show a strong decrease in power with increasing frequency, which is typical for the granulation signal, but no clear oscillation power excess. It is therefore difficult to decide a priori how many signal components are actually present in the data (i.e., statistically significant) and which of the measured components corresponds to granulation or something else. It can even happen that the PDS still includes some frequency-dependent instrumental signal, e.g., the spectral window or instrumental drifts. 

To avoid bias in the interpretation of the data we test various models (Eq.\,\ref{eq:globalmodel} with $i=1..n$, and each of them with and without a Gaussian) and let the global evidence $z$, as delivered by \textsc{MultiNest}, decide which model best represents the data. 

We start with the least complex model (i.e., one component and no Gaussian) and accept a more complex model (more components and/or a Gaussian) if its probability\footnote{In probability theory $p>0.9$ is considered as strong evidence \citep[e.g.,][]{Jeffreys1998}.} $p_i=z_i/\sum_j z_j > 0.9$. If a model is not accepted we add another degree of complexity and stop if the probability is again below 0.9. We find that the observed PDS of our sample of stars can be best reproduced by one to three components and find strong evidence for an oscillation power excess only in two stars, 39\,Cyg and $\sigma$\,Per. The best-fit models are shown in Fig.\,\ref{fig:stars0}--\ref{fig:stars3}. 

\begin{figure}[t]
	\begin{center}
	\includegraphics[width=0.5\textwidth]{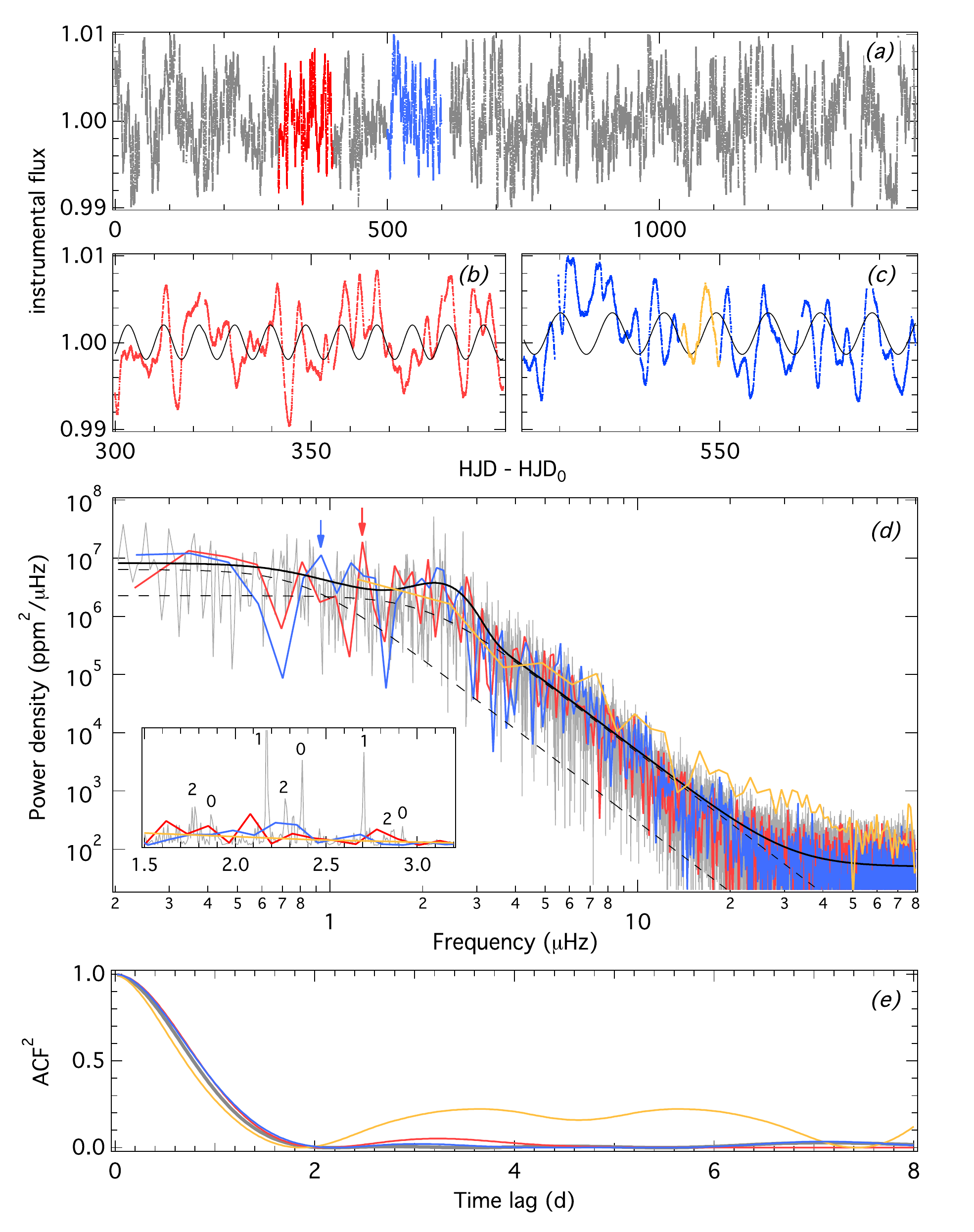}
	\caption{Full 1470 day-long \textit{Kepler} time series of KIC\,1431599 (grey; panel a) and 100 (red and blue) and 10 (orange) day-long subsets (panels b and c). Panel d shows the corresponding power density spectra along with a global fit (black line) consisting of two granulation components (dashed lines), a Gaussian, and instrumental white noise fitted to the full \textit{Kepler} time series (Eq.\,\ref{eq:globalmodel}) The largest amplitude peaks in the 100\,d PDS are marked with vertical arrows. The corresponding sinusoidal signal is overplotted to the respective time series. The inset shows the frequency range of the oscillation power excess with individual oscillation modes marked by their spherical degree. Panel e gives the autocorrelation functions of the various time series.}
	\label{fig:kepRG} 
	\end{center} 
\end{figure}

For several other targets clear variability is detected, but the BRITE time series are not long enough to properly detect an oscillation power excess (and resolve the oscillations modes within it). The interplay between intrisic variability time scales and observing length ($\Delta T$) is illustrated in Fig.\,\ref{fig:kepRG}, where we show the PDS of the full 1470 day-long \textit{Kepler} time series of KIC\,1431599 and for subsets of it. To resolve the star's oscillation power excess (i.e., its individual oscillation modes) around $\nu_\mathrm{max} \simeq 2.3$\mh\ requires considerably longer observations than the typical oscillation periods of about 5\,d. There is no clear rule for this but usually the mode pattern of a solar-type oscillator is resolved if the frequency resolution $df = 1/\Delta T$ is smaller than the frequency separation between $l=0$ and 2 modes $\delta\nu_{02}$. Taking into account the finite mode lifetimes and that $\delta\nu_{02}$ approximately scales as $0.04\nu_\mathrm{max}^{0.8}$ for red giants \citep[e.g.][]{Huber2010,kal2012} we define as a rule of thumb that $\Delta T \nu_\mathrm{max}^{0.8}\gg 25 $. In our sample of red giants only nine stars fullfill this criterion including 39\,Cyg and $\sigma$\,Per. In case of KIC\,1431599 this means that the star needs to be observed significantly ($\sim$3--5) longer than about 150 days to clearly detect and resolve the oscillation power excess. This is in fact shown in Fig.\,\ref{fig:kepRG}. While for the full \textit{Kepler} time series the power excess is clearly detectable with its individual oscillation modes resolved (see insert in panel \textit{d}), this is not the case for the two arbitrarily chosen 100 day-long subsets of the time series. Their PDS neither reveal individual modes  nor can the power excess be distinguished from the underlying granulation signal. On the other hand, the general behaviour of the granulation signal in the PDS remains unaffected from shortening the time series so that even for observations as short as 10 days one still gets the typical decrease in power with increasing frequency.

Another effect of the stochastic nature of the signal is that the dominant frequency (i.e., the frequency with the largest amplitude in the spectrum) in one data set can be quite different from that of other observations of the same star. This is because the probability that the largest-amplitude frequency in a random realisation of the stellar signal follows the limiting spectrum (i.e., of an infinitely long time series) of the granulation and oscillation signal, which is well approximated by our global fit. Since the limiting spectrum is almost flat below \num\ and only strongly decreases for higher frequencies, any frequency below \num\ has a similar chance to be the largest amplitude peak in an arbitrary data set. In case of KIC\,1431599 this means that we can find practically any period between $\Delta T$ and about 5\,d as the ``dominant signal'' of the time series. Or in other words, the seemingly dominant periodicity in the observations of a solar-type oscillator contain only limited information. Basically, little more can be gleaned than that the typical oscillation time scales are shorter.

In conclusion, for our sample of red giants this means that even though many of the present observations are too short to clearly detect an oscillation power excess, they are still long enough to extract valuable information from the granulation signal. Furthermore, several stars in our sample show clear variability with a seemingly dominant period when looking at their light curves. Even though these periods are of little help in our analysis, we measure them and list them in Tab.\,\ref{tab:results} for comparison with values found in the literature if their local (in a 10\mh\ box) signal-to-noise ratio is larger than four.

\begin{figure}
	\begin{center}
	\includegraphics[width=0.5\textwidth]{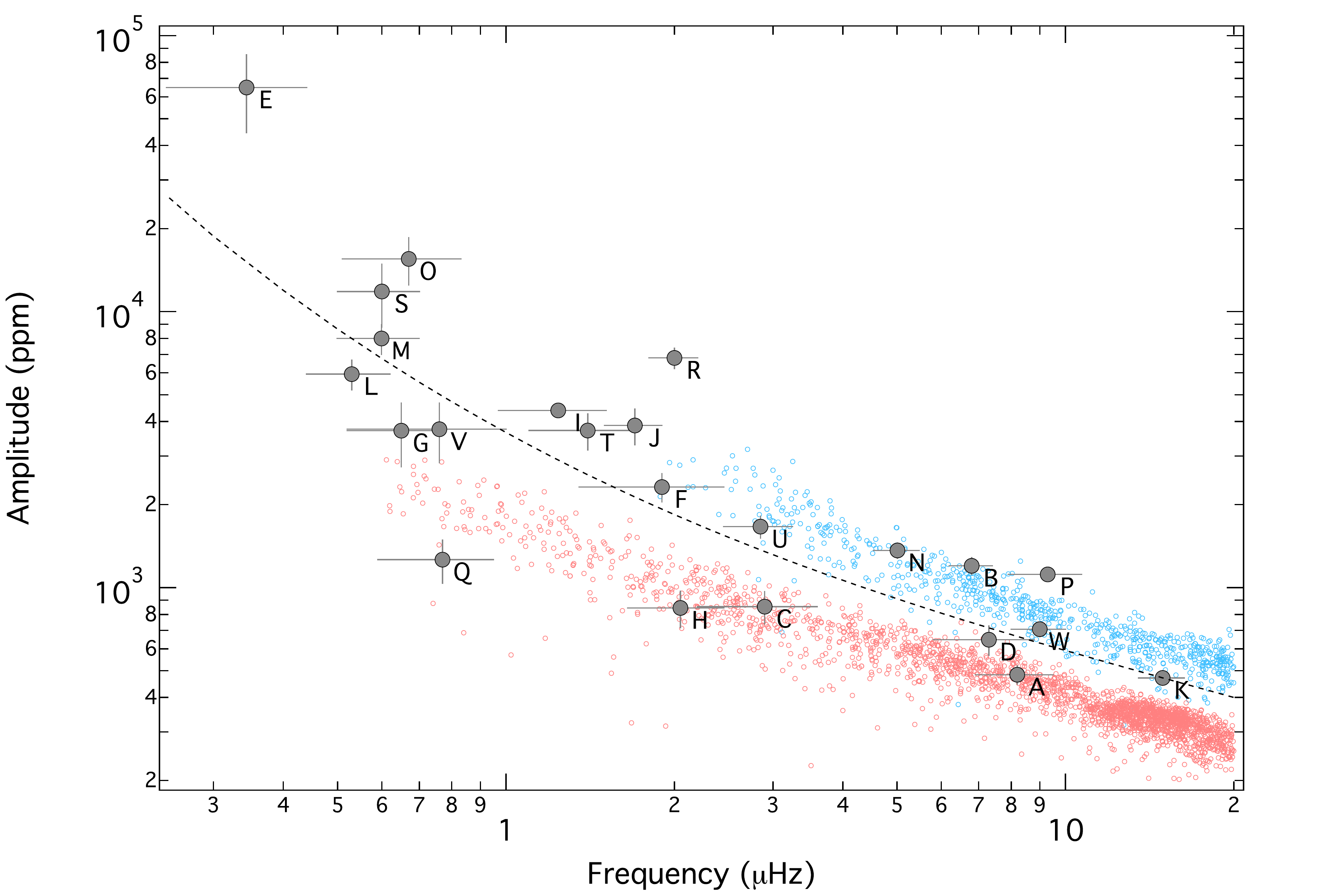}
	\caption{Low- (light-red circles) and high-frequency (light-blue circles) granulation components of a sample of red giant stars observed by \textit{Kepler}. The dashed line gives a fit to the mid point of the two components. Overplotted are the stars observed by BRITE-Constellation listed in Tab.\,\ref{tab:stars}.} 
	\label{fig:gran} 
	\end{center} 
\end{figure}

We now return to the actual analysis of the granulation signal in our sample of red giants. If we find more than one granulation component to be statistically significant in the PDS of a given star we concentrate on the component with the smallest frequency uncertainty. To identify to which granulation component it corresponds we compare the measured characteristic frequency and amplitude to those of a large sample of red giants observed with \textit{Kepler}. As is shown in Fig.\,\ref{fig:gran}, the LF and HF granulation components form two populations in the amplitude-frequency diagram. To discriminate between the branches we fit a $2^\mathrm{nd}$ order polynomial to the mid-point of the \textit{Kepler} measurements, which is determined by the average frequency and amplitude of the two granulation components. For a pair of parameters found for our target stars we can now identify the LF and HF component of the granulation signal. The measured characteristic granulation frequencies and \num\ are listed in Tab.\,\ref{tab:results}. Note that the amplitudes of both components are roughly the same \citep{kal2014}.

%--------------------------------------------------------------------
%--------------------------------------------------------------------

\begin{figure}
	\begin{center}
	\includegraphics[width=0.5\textwidth]{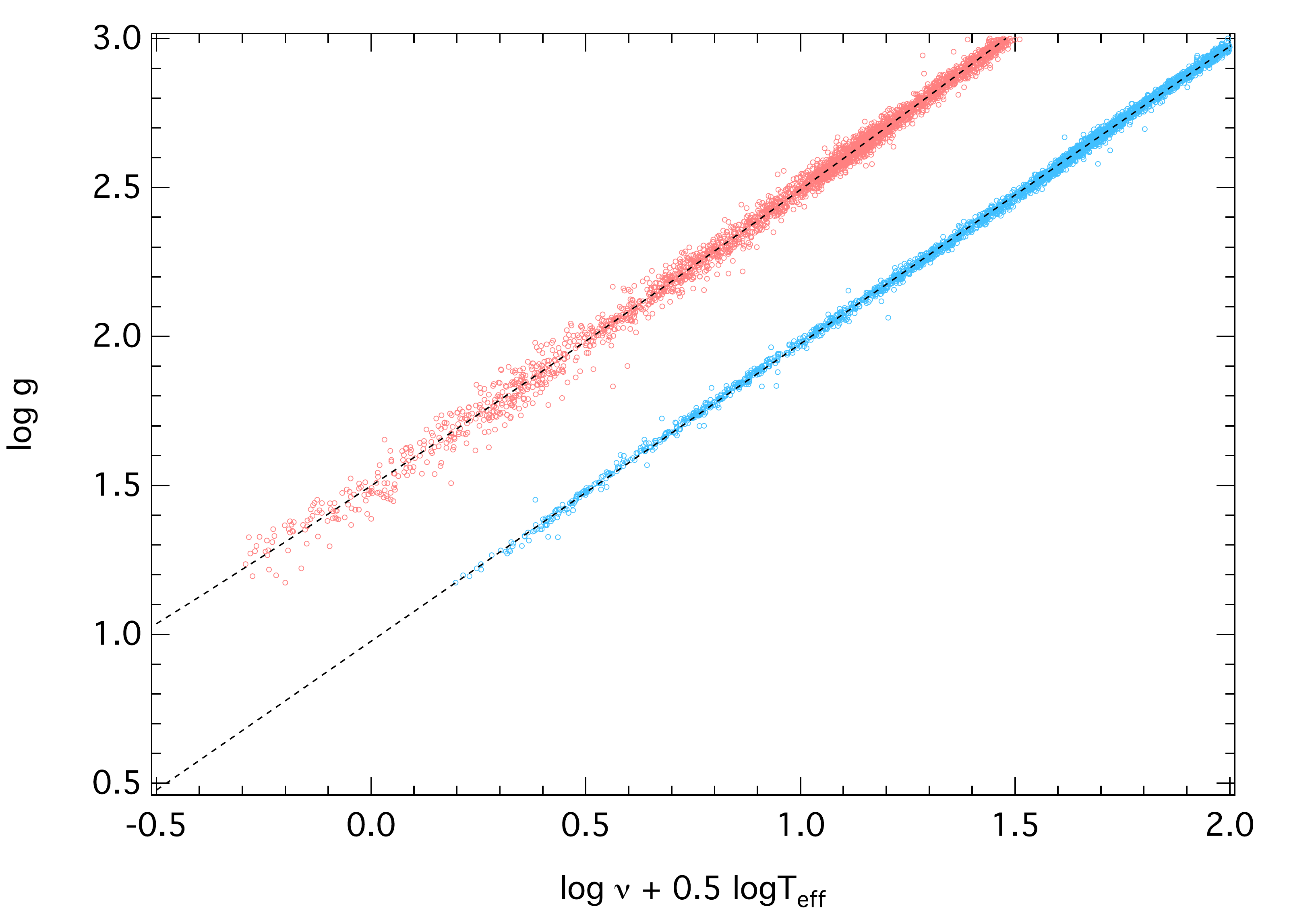}
	\caption{Surface gravity as a function of the frequency of the low- (light-red circles) and high-frequency (light-blue circles) granulation components of the sample of \textit{Kepler} stars. The dashed lines indicate fits to the data.} 
	\label{fig:loggfit} 
	\end{center} 
\end{figure}

With $\nu_\mathrm{low}$ and $\nu_\mathrm{high}$ determined, we can estimate the surface gravity \lg\ of the star. Even though we could directly define a scaling relation for \lg\ by combining Eq.\,\ref{eq:nugran} and \ref{eq:numax0}, \cite{kal2018} found evidence that the scaling of \num\ to \lg\ is more complex than is indicated by Eq.\,\ref{eq:numax0}. A better way is therefore to directly relate the measured granulation frequencies of the \textit{Kepler} sample with the stars' surface gravities, determined from the new seismic scaling relations. \cite{kal2018} have shown that \lg , determined from a full asteroseismic analysis, is accurate to better than 0.01 (about 2\% on a linear scale). The calibration is shown in Fig.\,\ref{fig:loggfit} yielding a relation: 
\begin{equation}	\label{eq:nuhigh}
\log g=0.977+0.998\mathcal{X}_\mathrm{high},
\end{equation}
for the HF component, and
\begin{equation}	\label{eq:nulow}
\log g=1.498+0.949\mathcal{X}_\mathrm{low}+0.045\mathcal{X}_\mathrm{low}^2
\end{equation}
for the LF component, with $\mathcal{X}_\mathrm{low,high}=\log \nu_\mathrm{low,high} + 0.5\log T_\mathrm{eff}$. The scatter around those fits is 0.012 ($\sim$2.8\%) and 0.02 ($\sim$4.5\%) for Eq.\,\ref{eq:nuhigh} and \ref{eq:nulow}, respectively, reflecting the typical uncertainties of the method. We finally apply these scaling relations to our sample of  stars and list the resulting \lg\ in Tab.\,\ref{tab:results}. In the case of 39\,Cyg and $\sigma$\,Per, the surface gravity is determined from \num\ according to \cite{kal2018} as 
\begin{equation}	\label{eq:numax}
\log g =  \log g\sun + (\nu_\mathrm{max}/\nu_\mathrm{max,\sun})^{1.0075\pm0.0021}\sqrt{T_\mathrm{eff}/5777\, \mbox{K}},
\end{equation}
where $\nu_\mathrm{max,\sun}=3140\pm5$\mh\ and $\log g\sun = 4.438$. We note that \lg\ values determined by $\nu_\mathrm{low}$ and $\nu_\mathrm{high}$ are consistent (but less accurate) with those based on \num .

%________________________________________________________________Tab. Results
\begin{table*}[t]
\begin{small}
\centering
\caption{Results for the target stars. $P$ is the approximate period of the signal with the largest amplitude in \bc\ data. $\nu_\mathrm{max}$, $\nu_\mathrm{low}$, and $\nu_\mathrm{high}$ are the frequency of the maximum oscillation power and the characteristic frequency of the low- and high-frequency component of the granulation signal, respectively. $\tau_\mathrm{ACF}$ corresponds to the width of sinc-function fitted to the autocorrelation function of the time series. The indices for \lg\ denote that the value is determined from a seismic or granulation measurement (S/G) or from the ACF time scale (ACF). Stellar radii ($R$) are listed as determined from measurements found in the literature (sources and methods are given in the notes). Stellar masses are derived directly ($M_\mathrm{direct}$) and via grid modelling ($M_\mathrm{grid}$). The last column gives the probability that a star is in the RGB phase of evolution (i.e., not to be in the RC phase).
\label{tab:results}}
\begin{tabular}{lr|ccccc|cc|c|cc|c}
\hline\hline
\noalign{\smallskip}
ID&Name &$P$& \num & $\nu_\mathrm{low}$ & $\nu_\mathrm{high}$&$\tau_\mathrm{ACF}$ &$\log g_\mathrm{S/G}$&$\log g_\mathrm{ACF}$&$R$&$M_\mathrm{direct}$&$M_\mathrm{grid}$&$p_\mathrm{RGB}$\\
&&[d]&\multicolumn{3}{c}{[\mh]}&[min]&\multicolumn{2}{c|}{[cm/s$^2$]}&[$R\sun$]&\multicolumn{2}{c|}{[$M\sun$]}&[\%]\\
\noalign{\smallskip}
\hline
\noalign{\smallskip}
A&$\eta$\,Cyg&56 &-&8.2$\pm$1.3&-&-&2.36$\pm$0.08&-&10.2$\pm$0.3$^c$&0.9$\pm$0.2&0.9$\pm$0.1&17\\

B&23\,Vul&-&-&-&6.8$\pm$0.6&795$\pm$27&1.75$\pm$0.04&1.74$\pm$0.02&31$\pm$2$^b$&1.9$\pm$0.3&1.9$\pm$0.3&29\\

C&39\,Cyg&-&9.4$\pm$0.7&2.9$\pm$0.7&-&550$\pm$16&1.83$\pm$0.03&1.91$\pm$0.02&25$\pm$1$^c$&1.9$\pm$0.1&1.8$\pm$0.2&31\\

D&$\varepsilon$\,Cyg&37&-&7.3$\pm$1.5&-&247$\pm$5&2.30$\pm$0.10&2.29$\pm$0.02&11.5$\pm$0.3$^d$&0.9$\pm$0.1&0.9$\pm$0.1&30\\

E&$\rho$\,Per&58&-&-&0.34$\pm$0.10&-&0.37$\pm$0.15&-&143$\pm$12$^h$&1.9$\pm$0.7&2.0$\pm$0.7&1\\

F&$\sigma$\,Per&36&4.5$\pm$0.6&1.9$\pm$0.6&-&1517$\pm$78&1.50$\pm$0.06&1.42$\pm$0.03&35$\pm$1$^c$&1.2$\pm$0.2&1.2$\pm$0.1&38\\

G&NS\,Pup&47&-&-&0.65$\pm$0.14&6028$\pm$38&0.70$\pm$0.10&0.73$\pm$0.02&213$\pm$24$^g$&9.1$\pm$2.1&8.0$\pm$0.9&18\\

H&$\beta$\,Pyx&-&-&2.0$\pm$0.4&-&-&1.75$\pm$0.09&-&24$\pm$2$^c$&1.2$\pm$0.3&1.3$\pm$0.3&56\\

I&g\,Car$^a$&12&-&-&1.2$\pm$0.3&3461$\pm$72&0.95$\pm$0.12&1.02$\pm$0.02&75$\pm$5$^b$&2.2$\pm$0.3&2.2$\pm$0.3&12\\

J&N\,Vel$^a$&9.0&-&1.7$\pm$0.2&-&2444$\pm$32&1.12$\pm$0.06&1.19$\pm$0.02&66$\pm$5$^h$&2.1$\pm$0.3&2.0$\pm$0.3&18\\

K&$\varepsilon$\,Sco&-&-&-&15$\pm$2&305$\pm$13&2.10$\pm$0.07&2.20$\pm$0.03&15.5$\pm$0.5$^c$&1.4$\pm$0.1&1.4$\pm$0.1&37\\

L&$\sigma$\,CMa&33&-&-&0.5$\pm$0.1&-&0.60$\pm$0.07&-&258$\pm$21$^h$&9.1$\pm$2.0&8.4$\pm$1.0&10\\

M&$\varepsilon$\,Mus&56&-&-&0.6$\pm$0.1&7737$\pm$39&0.62$\pm$0.09&0.60$\pm$0.02&116$\pm$9$^h$&2.0$\pm$0.3&2.1$\pm$0.3&5\\

N&$\varepsilon$\,Cru&-&-&-&5.0$\pm$0.5&1029$\pm$48&1.60$\pm$0.05&1.61$\pm$0.03&31$\pm$2$^c$&1.4$\pm$0.2&1.5$\pm$0.2&37\\

O&$\gamma$\,Cru&22&-&-&0.7$\pm$0.2&5702$\pm$36&0.71$\pm$0.14&0.75$\pm$0.02&84$\pm$7$^e$&1.5$\pm$0.2&1.5$\pm$0.3&29\\

P&G\,Sco&82&-&-&9.3$\pm$1.4&533$\pm$45&1.88$\pm$0.07&1.93$\pm$0.05&20$\pm$1$^c$&1.2$\pm$0.2&1.2$\pm$0.2&24\\

Q&$\gamma^2$\,Sgr&44&-&-&16$\pm$2&371$\pm$18&2.13$\pm$0.05&2.10$\pm$0.03&11.8$\pm$0.6$^c$&0.6$\pm$0.1&0.7$\pm$0.1&43\\

R&$\eta$\,Sgr&8.6&-&-&1.8$\pm$0.2&2717$\pm$17&1.13$\pm$0.05&1.12$\pm$0.02&66$\pm$12$^c$&2.1$\pm$0.7&2.0$\pm$0.7&27\\

S&HD\,19349&46&-&-&0.6$\pm$0.1&5496$\pm$58&0.65$\pm$0.09&0.77$\pm$0.02&87$\pm$8$^b$&1.6$\pm$0.3&1.7$\pm$0.3&30\\

T&$\alpha$\,Cet&14&-&-&1.4$\pm$0.3&3366$\pm$95&1.04$\pm$0.09&1.03$\pm$0.02&89$\pm$5$^f$&3.8$\pm$0.5&3.9$\pm$0.5&3\\

U&V337\,Car&-&-&-&2.8$\pm$0.4&2546$\pm$62&1.36$\pm$0.06&1.17$\pm$0.02&128$\pm$12$^c$&9.0$\pm$1.6&7.3$\pm$0.9&57\\

V&GZ\,Vel&53&-&-&0.8$\pm$0.2&5252$\pm$54&0.78$\pm$0.12&0.80$\pm$0.02&141$\pm$11$^c$&4.6$\pm$0.7&4.5$\pm$0.7&0\\

W&12\,Mus&62&-&-&9.0$\pm$1&498$\pm$30&1.89$\pm$0.05&1.96$\pm$0.03&16.6$\pm$1$^b$&1.0$\pm$0.3&1.1$\pm$0.3&31\\
\hline
\end{tabular}
\tablefoot{
based on:
\tablefoottext{a}{2015 measurements}--
\tablefoottext{b}{Gaia DR2 Catalogue}--
\tablefoottext{c}{interferometric angular diameter \citep{Richichi2005} and Gaia DR2 parallax}--
\tablefoottext{d}{\cite{Piau2011}}--
\tablefoottext{e}{\cite{Ireland2004}}--
\tablefoottext{f}{\cite{Wittkowski2006}}--
\tablefoottext{g}{\cite{McDonald2017}}--
\tablefoottext{h}{$R^2=L/T_\mathrm{eff}^4$}
}
\end{small}
\end{table*}
%__________________________________________________________________

%--------------------------------------------------------------------
%--------------------------------------------------------------------

The uncertainties in our \lg\ result from the combined uncertainties of the observables ($\nu_\mathrm{low,high}$ and $T_\mathrm{eff}$) and the calibration and range from 0.03 ($\sim$7\%) to 0.1 ($\sim$23\%), which is still quite large for a seismology-based method. Even though the \lg\ estimates are roughly consistent with literature values, it is not guaranteed that we do not misinterpret signal from rotation, magnetic activity, or even from the instrument as granulation signal, which would seriously affect our analysis. A possibility to verify this is given by the following method.

%--------------------------------------------------------------------
%--------------------------------------------------------------------

\subsection{The autocorrelation method}
There are various other methods that make use of the granulation signal to estimate \lg\ of stars in the \textit{Kepler} field \citep{Hekker2012,Bastien2016,Bugnet2018,Pande2018}. They all rely on the amplitude of the brightness variations due to granulation measured in one or more given frequency bands. However, such amplitudes not only depend on the surface gravity of the star but also on other (gravity-independent) properties, like rotation and activity. Furthermore, they are calibrated to the \textit{Kepler} passband and instrumental noise and can therefore not directly be applied to observations with other instruments. A more flexible and accurate approach for a \lg\ determination has been introduced by \cite{kal2016} who have shown that one can estimate \lg\ solely from the typical timescale of the combined granulation and oscillation signal, which they derive from the autocorrelation function (ACF) of the time series. The ACF of a stochastic signal (granulation and solar-type oscillations) follows a sinc-function and its width defines a typical time scale $\tau_\mathrm{ACF}$, which scales with the surface gravity of the star. The method gives good results even for short time series (see Fig.\,\ref{fig:kepRG}) and data where the oscillations are hidden in the noise. Their calibration for a large sample of \textit{Kepler} stars, covering the entire region from the main sequence up to the giant branch, results in
\begin{equation}	\label{eq:tauacf}
\log g = 4.766 - 0.962 \log (\tau_\mathrm{ACF}) - 0.026 \log^2 (\tau_\mathrm{ACF}),
\end{equation}
which is accurate to about 0.017 (or about 4\%). 

A problem of this approach is that it is calibrated with  \lg\ values from seismology and, hence, relies on the classical $g \propto \nu_\mathrm{max}\sqrt{T_\mathrm{eff}}$ scaling. As mentioned above, \cite{kal2018} found that $g$ better scales according to Eq.\,\ref{eq:numax}, which also influences the $\tau_\mathrm{ACF}$ calibration, especially for stars with low \lg . Instead of re-calibrating Eq.\,\ref{eq:tauacf} (which is beyond the scope of this analysis), we add a correction term
\begin{equation}
\Delta = 0.0075\pm0.0021 [\log g - \log g\sun - 0.5\log(T_\mathrm{eff}/5777\, \mbox{K})],
\end{equation}
to Eq.\,\ref{eq:tauacf}, which accounts to first approximation for the ``non-linearity'' of the \num\ scaling relation.

The autocorrelation functions of our target stars are shown in Fig.\,\ref{fig:stars0}--\ref{fig:stars3} along with the best-fit squared sinc-functions. The resulting $\tau_\mathrm{ACF}$ and \lg\ values are given in Tab.\,\ref{tab:results} (see also Fig.\,\ref{fig:hrd}), where the uncertainties for the latter cover again all error sources, from the observables to the calibration itself, and amount typically to about 0.02 (or 4.5\%).

We find generally good agreement for \lg\ values derived with the ACF method and from granulation time scales, with the former giving more accurate results. Only for the two stars NS\,Pup and GZ\,Vel does a significant difference exist. According to Fig.\,\ref{fig:gran} the observed signal is identified as the low-frequency component of the granulation signal, indicating a \lg\ of 1.24 and 1.32 for NS\,Pup and GZ\,Vel, respectively. These values are significantly different from those based on $\tau_\mathrm{ACF}$. However, if we assume the observed signal to be due to the high-frequency granulation component (i.e., using Eq.\,\ref{eq:nuhigh} instead of Eq.\,\ref{eq:nulow} to determine \lg ) we get good agreement (see Tab.\,\ref{tab:results}). In this case, the granulation amplitude would have been underestimated. In fact, this also happens for some \textit{Kepler} stars as is indicated in Fig.\,\ref{fig:gran} with blue dots below the dashed line. Possible explanations are strong magnetic fields suppressing the granulation amplitude or dilution of the observed light by another star (e.g., from a binary companion). 

\section{Stellar mass and evolutionary stage}

Stellar masses are only known for about half of our target stars and those are often questionable as they mostly result from a comparison of spectroscopic measurements with stellar evolution models. Consequently, they are highly model dependent, so that any improvements based on actual observations are valuable for future analyses. 

Given the surface gravities from above we could directly calculate the mass as,
\begin{equation}
M=gR^2, 
\end{equation}
(all in solar units) if there is a reliable estimate for the radius $R$ available. Since all stars in our sample are bright stars that are quite close and on the giant branch (i.e., large objects) we find in fact precise measurements of the interferometric angular diameter $\Theta$ in the Catalog of High Angular Resolution Measurements (CHARM) of \cite{Richichi2005} for 11 stars in our sample. With these measurements and with the Gaia DR2 parallaxes $\pi$ \citep{GaiaDR2} we can directly determine the radius (in solar units) as,
\begin{equation}
R = 107.548 \frac{\Theta}{\pi}.
\end{equation}
For three stars we use the radius estimate from the Gaia DR2 catalog. For four stars no entries in the Gaia catalog are available yet but we find radius estimates in the literature. Only for $\rho$\,Per, $\sigma$\,CMa, and $\varepsilon$\,Mus no such measurements are available, so that we determine the radius according to $R^2=L/T_\mathrm{eff}^4$ (all in solar units), with the effective temperatures and luminosities listed in Tab.\,\ref{tab:stars}. If available, we use \lg\ from the ACF method to determine the mass.

A special case is the star N\,Vel, for which the CHARM and Gaia measurements result in a radius of about 33\,$R$\sun , which in turn would yield a mass as small as 0.5\,$M$\sun . In fact, the interferometric radius is incompatible with the value determined from $L/T_\mathrm{eff}^4$ so that there seems to be a problem with the interferometric angular diameter. We therefore adopt the radius that results from the stars' luminosity and effective temperature. The corresponding radii and resulting basically model-independent masses are listed in Tab.\,\ref{tab:results} with typical uncertainties of 3--9\% in radius and 10--20\% in mass.

\begin{figure}
	\begin{center}
	\includegraphics[width=0.5\textwidth]{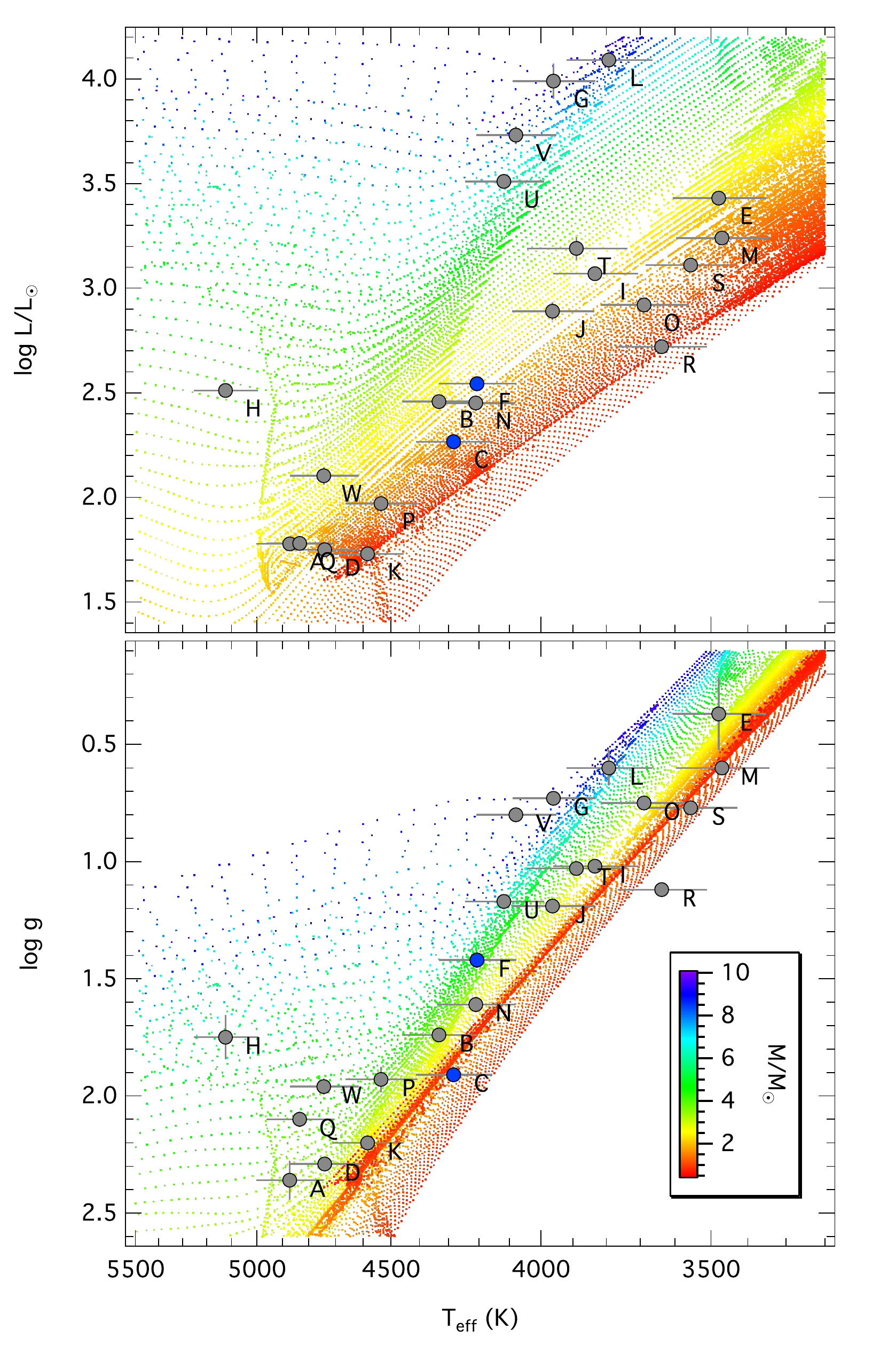}
	\caption{Hertzsprung-Russell (top) and Kiel-diagram (bottom) including the red giants observed by \bc\ (grey-filled circles). The small dots show MIST stellar evolution models for solar composition with the mass colour-coded. Blue-filled circles mark stars for which solar-type oscialltions have been found in the \bc\ data.} 
	\label{fig:hrd} 
	\end{center} 
\end{figure}

Another method to derive the stellar mass is to compare the \lg , $T_\mathrm{eff}$, and $R$ measurements with those of a grid of stellar models. We adapt the grid-modelling approach from \cite{kal2010} using \lg , $T_\mathrm{eff}$, and $R$ as input instead of the original global seismic parameters. The model grid is extracted from the non-rotating version of the MESA\footnote{Modules for Experiments in Stellar Astrophysics \citep{Paxton2011}} Isochrones \& Stellar Tracks \citep[MIST\footnote{\href{url}{http://waps.cfa.harvard.edu/MIST/index.html}};][]{Dotter2016,Choi2016}. The grid covers about 900\,000 models from the main-sequence phase to the asymptotic giant branch with initial masses from 0.1 to 10\,$M\sun$, and initial chemical compositions of [Fe/H] = $-$4.0 to 0.5. The Bayesian comparison between observed and model parameters accounts for all observational uncertainties. The solar-composition part of the grid is shown in Fig.\,\ref{fig:hrd}. The resulting mass estimates are listed in Tab.\,\ref{tab:results}, which are generally in good agreement with the directly determined masses. 

The main reason for the grid-modelling, however, is that we can evaluate the evolutionary states of our target stars in a statistical sense. The model grid consists of isochrones with time steps of equal length. Consequently, we will find only few grid points for parameter areas with fast stellar evolution. If we now add up the model probabilities according to their evolutionary state we can assign each star a probability to be in a given stage of stellar evolution. In practice, we compute the probability $p_\mathrm{RGB}$ that a star is a RGB star, or vice-versa that the star is a post-RGB star with the probability of $1-p_\mathrm{RGB}$ \citep[see also][]{Hekker2017b}. In fact, with high confidence ($p_\mathrm{RGB} < 9\%$, i.e. the odds ratio is better than about 10:1) we can only determine the evolutionary state of five stars ($\rho$\,Per, N\,Vel, $\varepsilon$\,Mus, $\alpha$\,Cet, and GZ\,Vel) and as post-RGB’s. For five more stars ($\eta$\,Cyg, NS\,Pup, g\,Car, $\sigma$\,CMa, and G\,Sco), the probability contrast ($p_\mathrm{RGB} < 25\%$, i.e. the odds ratio is better than 3:1) is high enough to claim that they are post-RGB’s, but for the remaining stars we can only make a tentative statement about their evolutionary state or provide no information at all if  $p_\mathrm{RGB}$ is around 50\%.

%--------------------------------------------------------------------
%--------------------------------------------------------------------

\subsection{Comparison with literature values}
\begin{figure}
	\begin{center}
	\includegraphics[width=0.5\textwidth]{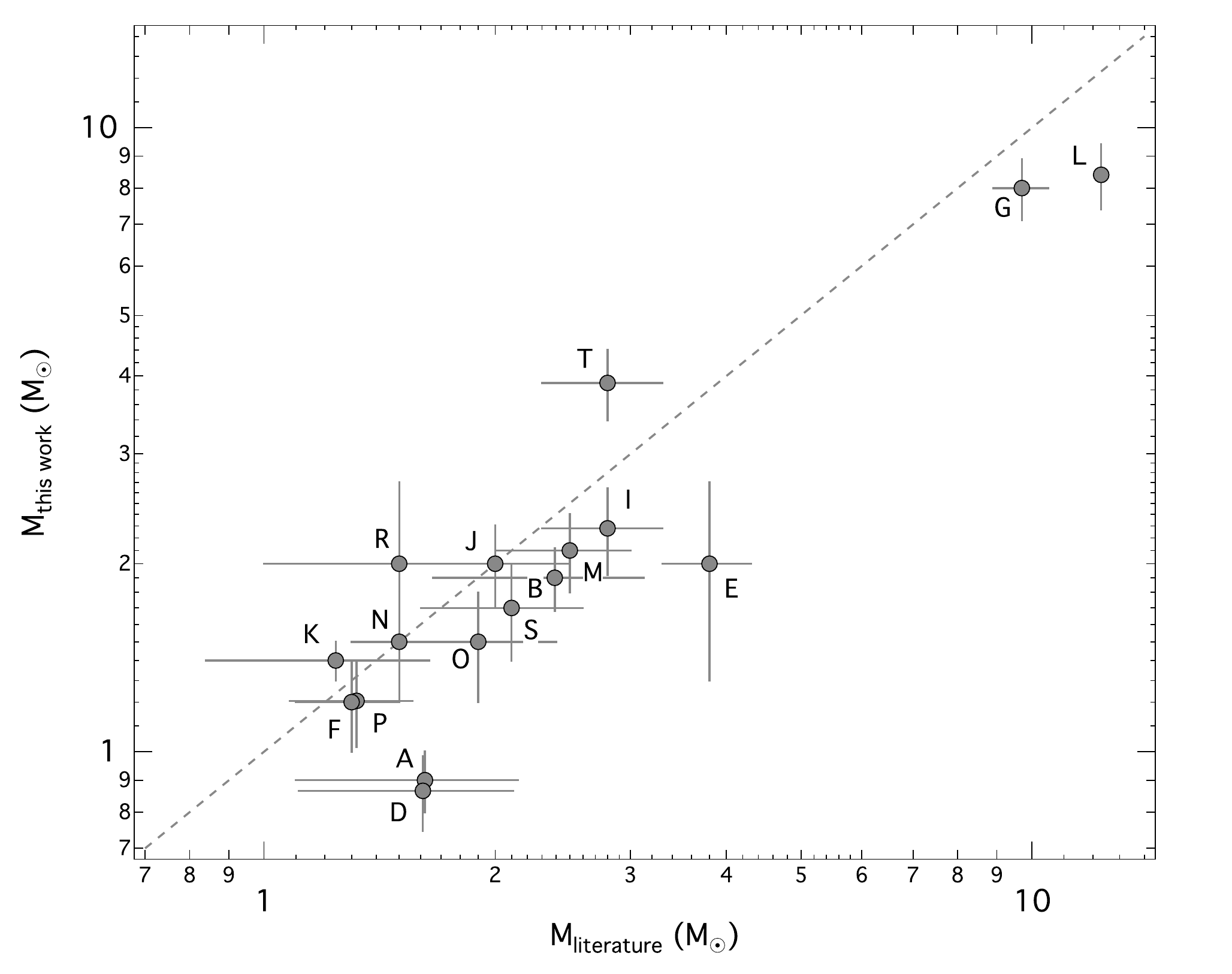}
	\caption{Comparison between masses determined in this work and found in the literature.} 
	\label{fig:m_comp} 
	\end{center} 
\end{figure}

Unless a star is gravitationally bound in a binary system, reliable mass estimates are difficult to achieve and mostly rely on the comparison of some observed parameters (\teff , \lg , $L$, etc.) with stellar evolution models. We find mass estimates in the literature for 17 stars of our sample. They are discussed in the following along with known variability time scales:
\begin{itemize}
\item (A) \textbf{$\eta$\,Cyg}: Isochrone fitting by \cite{Luck2015} yields a mass ranging from 1.10 to 2.13\,$M$\sun . 
\item (B) \textbf{23\,Vul}: Based on a 25.3\,yr binary orbit \cite{Malkov2012} determined a dynamical mass of $2.39\pm0.73$\,$M$\sun .
\item (D) \textbf{$\varepsilon$\,Cyg}: Based on model atmospheres and stellar evolutionary tracks \cite{Fuhrmann2017} claim a mass of 1.61\,$M$\sun , for which we expect the uncertainties to be on the order of $\pm$0.5\,$M$\sun . 
\item (E) \textbf{$\rho$\,Per}: \cite{Dumm1998} found a mass of 3.8\,$M$\sun\ from comparison of a Hipparcos parallax-based luminosity and spectroscopic effective temperature with a grid of stellar models. We assume an uncertainty of $\pm$0.5\,$M$\sun . The GCVS lists a period of 50\,d, which is consistent with the dominant period of 58\,d in the BRITE observations.
\item (F) \textbf{$\sigma$\,Per}: Isochrone fitting by \cite{Reffert2015} gives $1.32\pm0.24$\,$M$\sun .
\item (G) \textbf{NS\,Pup}: Comparison of atmospheric parameters with various evolutionary models results in a median mass of $9.7\pm0.8$\,$M$\sun\ \citep{Tetzlaff2011}.
\item (I) \textbf{g\,Car}: $2.8\pm0.5$\,$M$\sun\ from \cite{Dumm1998}. The 187\,d period given by \cite{Price2010} is not detectable in the 61 day-long BRITE observations, in which we find a period of about 12\,d.
\item (J) \textbf{N\,Vel}: \cite{Gondoin1999} found a mass of about 2\,$M$\sun\ based on comparison with evolutionary models, for which we expect the uncertainties to be on the order of $\pm$0.5\,$M$\sun .  
\item (K) \textbf{$\varepsilon$\,Sco}: Based on model atmospheres and stellar evolutionary tracks \cite{Fuhrmann2017} claim a mass of 1.24\,$M$\sun, for which we expect the uncertainties to be on the order of $\pm$0.4\,$M$\sun . 
\item (L) \textbf{$\sigma$\,CMa}: \cite{Tetzlaff2011} find a median mass of $12.3\pm0.1$\,$M$\sun\ from comparison of atmospheric parameters with different evolutionary models. 
\item (M) \textbf{$\varepsilon$\,Mus}: $2.5\pm0.5$\,$M$\sun\ from \cite{Dumm1998}. The GCVS gives a period of 40\,d and \cite{Tabur2009} list several periods between 32 and 63\,d, which is consistent with the 58\,d period found in the BRITE data.
\item (N) \textbf{$\varepsilon$\,Cru}: \cite{Jofre2015} performed a detailed analysis of high-resolution spectroscopic observations and found a mass of $1.5\pm0.2$\,$M$\sun\ based on comparison with stellar evolution models. 
\item (O) \textbf{$\gamma$\,Cru}: $1.9\pm0.5$\,$M$\sun\ from \cite{Dumm1998}. \cite{Tabur2009} lists periods between 12 and 83\,d, which is consistent with the 22\,d period found in BRITE observations.
\item (P) \textbf{G\,Sco}: \cite{Stello2008} determined a mass of $1.4\pm0.2$\,$M$\sun\ from seismic observations with the WIRE satellite, which reduces to $1.3\pm0.2$\,$M$\sun\ when using the Gaia parallax and the non-linear scaling from \cite{kal2018}. 
\item (R) \textbf{$\eta$\,Sgr}: $1.5\pm0.5$\,$M$\sun\ from \cite{Dumm1998}.
\item (S) \textbf{HD\,19349}: $2.1\pm0.5$\,$M$\sun\ from \cite{Dumm1998}.
\item (T) \textbf{$\alpha$\,Cet}: $2.8\pm0.5$\,$M$\sun\ from \cite{Dumm1998}.
\end{itemize}
In Fig.\,\ref{fig:m_comp} we confront our mass estimates with those from the literature and find good agreement for several stars, especially for the stars 23\,Vul (B) and G\,Sco (P) for which model-independent mass estimates are available in the literature. For the remaining stars, inconsistency can be explained by the mostly unconsidered systematic uncertainties in the literature values, like the degeneracy between \teff , \lg , and [Fe/H]. Furthermore, the observed atmospheric parameters of a star are usually compared to those of a single set of stellar evolution models not considering the systematic errors in the models that arise from, e.g., the insufficient physics used to produce the models. Even more, simply using models computed with similar input parameters and physics but with a different stellar evolution code can result in significantly different results. In this context, \cite{Stancliffe2016} have shown that solar-calibrated evolutionary tracks from various stellar evolution codes differ by up to 150\,K when leaving the main sequence, which translates into a systematic mass uncertainty of about 5\% on the main sequence and some 10\% on the giant branch. On top of that, the often unknown initial hydrogen and metal abundances produce additional uncertainties as well as uncertain input physics, like the treatment of convection, uncertain nuclear reaction rates and opacities, rotational mixing, etc, do. Furthermore, several stars in our sample are already in a luminosity regime where mass loss can play a substantial role, further complicating the mass determination via stellar models. In conclusion, the mass of a red giant determined from stellar evolution models might serve as guidance to distinguish between a high-mass star and a low-mass star but should be taken with care beyond that. We are aware that our grid-modelling approach suffers from the same deficiencies but note that the masses determined from the grid modelling are fully consistent with those from the model-independent method and should therefore be preferred over the literature values.

%--------------------------------------------------------------------
%--------------------------------------------------------------------

\section{Period-luminosity relation}
Thanks to large ground-based microlensing surveys such as MACHO and OGLE, small-amplitude oscillations have been observed in thousands of luminous red giants \citep[e.g.][and references therein]{Wood1999,Wray2004,Soszynski2007,Tabur2010,Soszy2013}. However, the nature of these oscillations has been questioned for some time. In particular, it is debated whether they are self-excited pulsations or stochastically excited modes \citep[e.g.][]{Bedding2003,JCD2001,Dziembowski2010}. Most of the results of the microlensing surveys are expressed as period-luminosity relations found for different sequences of oscillation modes \citep[e.g.][]{Soszynski2007}. These sequences smoothly connect with the pattern of radial and non-radial modes observed in \textit{Kepler} red giants \citep[][]{Mosser2013,Stello2014} with the same scaling relation that links the peak frequency to the frequency separation. It is therefore quite likely that the oscillations observed in semi-regular and even Mira-type variables are caused by stochastic excitation, similar to those found in the Sun.

\begin{figure*}
	\begin{center}
	\includegraphics[width=1.0\textwidth]{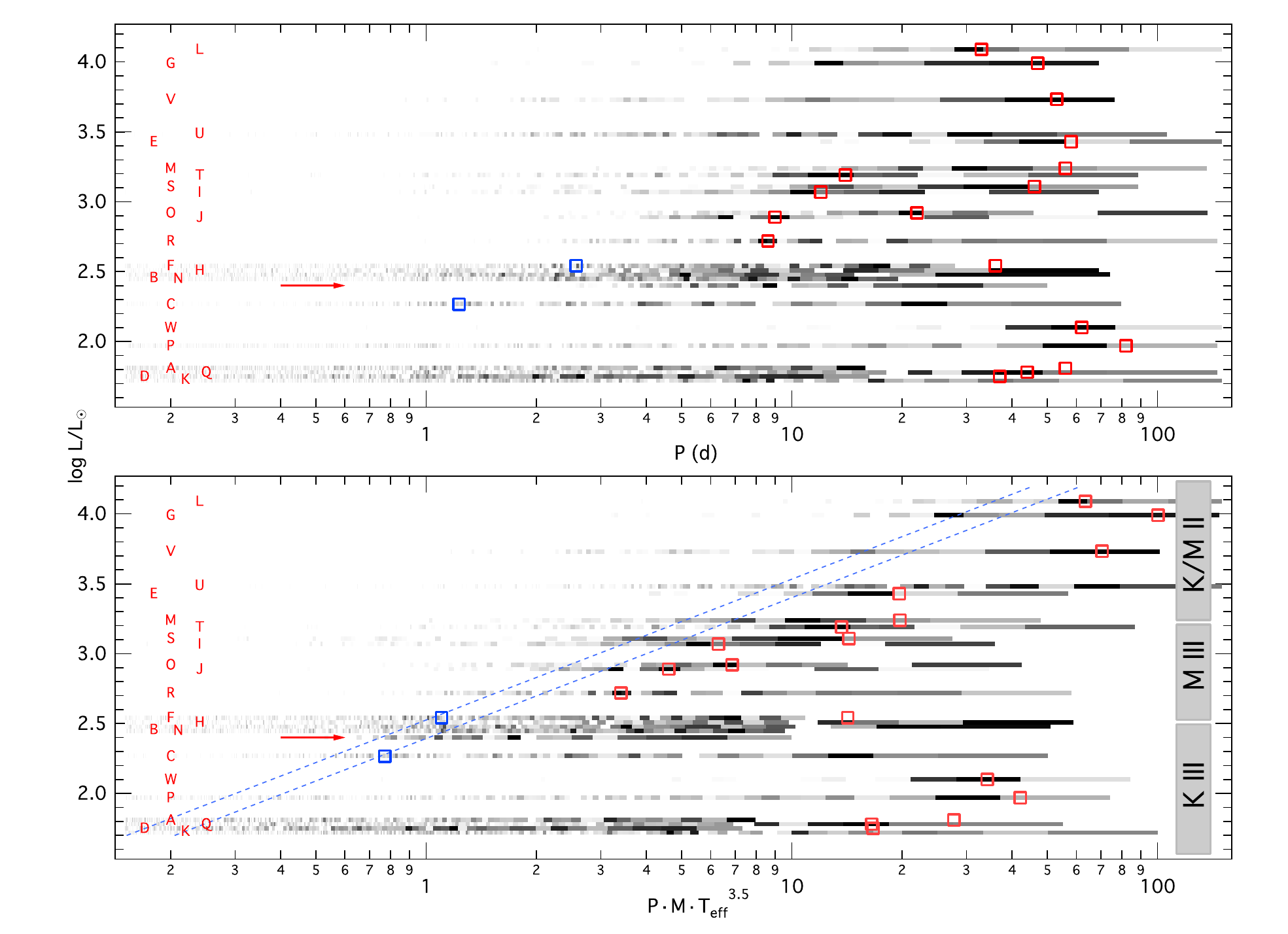}
	\caption{Observed power spectra as a function of period (top) and scaled period (bottom) of our sample stars. The stars are sorted according to their luminosity and displayed as horizontal band with the level of power indicated by the gray scale. The spectrum of KIC\,1431599 (from Fig.\,\ref{fig:kepRG}) is indicated by a red arrow. Open red squares correspond to the dominant periods in the BRITE data and open blue squares indicate our measurements of \num . The period range were to expect oscillations is marked with dashed lines. Typical spectral types and luminoisty classes are given on the right-hand side.} 
	\label{fig:plr} 
	\end{center} 
\end{figure*}

Even though the limited length of our \bc\ observations did not allow us to resolve individual oscillation modes, we can still use the period-luminosity relation to verify the origin of the observed signal. This is shown in Fig.\,\ref{fig:plr}, where the observed PDS are sorted according to the star's luminosity and displayed as horizontal bands with the level of power indicated by the gray scale (top panel). There is already a clear trend that for increasing luminosity the variability timescales increase. This trend is, however, smeared by stellar mass in the sense of that more massive stars show their variability at lower timescales than less massive stars with the same luminosity. Such a correlation is in fact well approximated by Eq.\,\ref{eq:numax0}, which can be written as, $\nu_\mathrm{max} = M T_\mathrm{eff}^{3.5} / L$ (all in solar units). If we now scale the period axis of the observed PDS with $MT_\mathrm{eff}^{3.5}$ we can expect a linear relation between the variability time scales and $L$, which is in fact shown in the bottom panel of Fig.\,\ref{fig:plr}. Practically all stars show their strongest signal above the period range at which oscillations are expected. This confirms that we do not misinterpret poorly resolved oscillations as granulation signal and that the granulation timescales in luminous red giants follow the same basic rules as in MS stars. Furthermore, it confirms that the dominant periods measured in the BRITE data are basically randomly distributed so that their information content is rather limited (other than that they are longer than the typical oscillation time scales). In fact, we find only two stars ($\sigma$\,CMa and g\,Car) for which the dominant period comes close to the period range where oscillations are expected so that the visible variability in their light curves might rather be due to oscillations than to granulation.

%--------------------------------------------------------------------
%--------------------------------------------------------------------

\section{Summary and conclusion}
In recent years seismology of red giants has grown to become an important field in stellar astrophysics, providing the unique opportunity to probe the interior structure of evolved stars. Apart from this, seismic scaling relations have become a valuable method to determine the mass and radius of stars with a convective envelope. Such studies are, however, limited to relatively faint stars accessible to the long-term high-precision observations with CoRoT and \textit{Kepler} with the exception of more recent observations of a few naked-eye stars in the ecliptic plane using special-aperture photometry with K2 \citep{White2017}. Bright stars, on the other hand, with plenty of additional information available (from spectroscopy, interferometry, etc.) are seismically less well constrained, which is even truer for red giants requiring long-term time series observations.

In this paper we present the \bc\ observations of a sample of red giant stars in which we find a clear granulation and/or oscillation signal. The sample consists of 23 stars distributed all over the sky with V magnitudes ranging between 1.6 and 5.0 and covers low-mass red clump stars to high-mass red giants. Each star has been observed almost continuously by at least one of five BRITE satellites for up to 173\,d.

Even though plenty of information is available in the literature about this sample, for none of the stars is either \lg\ or the mass sufficiently well known as one might expect for such bright stars. We therefore use the granulation and/or oscillation time scales measured from the \bc\ observations to determine model-independent estimates of \lg\ with two different methods. Using precise radii from the literature (mostly from interferometric angular diameters and Gaia parallaxes) we can then determine the mass of the stars from our \lg\ measurements.

We confirm that the \bc\ data are dominated by the granulation signal and that the corresponding typical time scales follow the same scalings as the typical oscialltion periods even for stars high up the giant branch (G to M-type Supergiants). We measure the individual time scales in Fourier space as well as in the time domain and use corresponding scaling relations to convert these time scales into surface gravities. To account for the non-linearity in the $\nu_\mathrm{max}-g$ relation recently found by \cite{kal2018} we first re-calibrate the corresponding scalings using measurements from a large sample of red giants observed with \textit{Kepler}.

\cite{kal2014} have shown that in Fourier space the granulation signal consists of at least two components, each comprising a typical time scale and amplitude, which tightly scale with the surface gravity of the star. In noisy data, however, or in data of insufficient length or sampling, one can often detect only one component, which is then difficult to assign. Using measurements of the \textit{Kepler} giants we find an empirical relation between the granulation time scale and amplitude that allows us to evaluate whether a measured set of parameters originates from the LF or HF component of the granulation signal in order to apply the correct \lg\ scaling. 

We can determine \lg\ for all but four stars of our sample from the power density spectrum as well as from the ACF of the time series and find them to agree within the uncertainties, where the ACF method is typically more accurate.

Based on these \lg\ measurements and stellar radii from the literature we can then directly determine the masses of our sample stars. Comparison with parameters from a large grid of stellar models also allows us to statistically evaluate the evolutionary state of the individual stars. We find that five stars are likely and five stars are very likely post-RGB stars. For the remaining stars the probability contrast is not high enough to make more than a tentative statement.

The stellar masses presented here range from about 0.7 to more than 8\,$M$\sun\ and have formal uncertainties of about 10-20\%, which covers the observational errors as well as the known uncertainties of the used scaling relations. One might question that our simple scaling relations hold from low-mass giants with about 10\,$R$\sun\ to high-mass giants with more than 200\,$R$\sun . Lacking independent and reliable mass estimates this is difficult to verify. Even though there might still be some unknown systematics in the scaling relations, they appear to be at least good enough to disentangle low-mass stars from high-mass stars and given the large unknown uncertainties in the model-based masses we prefer our model-independent estimates. 

Out of the 23 stars in our sample only two (39\,Cyg and $\sigma$\,Per) show a significant oscillation power excess, from which we can measure \num . Even though their \num\ values differ by more than a factor of two, they appear to have a very similar effective temperature of around 4200\,K (i.e., spectral type of around K3). We think that this temperature represents some kind of a ``sweet spot'' for observing solar-type oscillations with \bc . For cooler stars the time base of the observations is not long enough to clearly resolve the signal and for hotter stars the oscillation amplitudes are too small to be detected by the BRITE instruments. In fact, there are two other stars (23\,Vul and $\varepsilon$\,Cru) in this temperature range with no detected oscillation signal. For 23\,Vul the oscillation signal seems to interfere with the instrumental 1\,d$^{-1}$ signal and for $\varepsilon$\,Cru there is some signal at about 4\mh , which is, however, not statistically significant. We further note that another star in this spectral range ($\alpha$\,Tau) shows clear solar-type oscillations in \bc\ data. The observations are, however, not yet finished, which is why the star is excluded from the present analysis. The data will be published in the near future.

\begin{acknowledgements}
TK, WW, and KZ are grateful for funding via the Austrian Space Application Programme (ASAP) of the Austrian Research Promotion Agency (FFG) and BMVIT. PGB acknowledges the support of the Spanish Ministry of Economy and Competitiveness (MINECO) under the programme ’Juan de la Cierva incorporacion’ (IJCI-2015-26034). D.H. acknowledges support by the National Aeronautics and Space Administration under Grant NNX14AB92G issued through the Kepler Participating Scientist Program and support by the National Science Foundation (AST-1717000). The research leading to the presented results has received funding from the European Research Council under the European Community’s Seventh Framework Programme (FP7/2007-2013) / ERC grant agreement no 338251 (StellarAges). GH acknowledges support from the Polish National Science Center (NCN), grant no. 2015/18/A/ST9/00578. AFJM acknowledges financial aid from NSERC (Canada) and FQRNT (Quebec). APi acknowledges support from the NCN grant 2016/21/B/ST9/01126. APo was responsible for image processing and automatation of photometric routines for the data registered by BRITE-nanosatellite constellation, and was supported by NCN grant 2016/21/D/ST9/00656. GAW acknowledges Discovery Grant support from the Natural Sciences and Engineering Research Council (NSERC) of Canada.
\end{acknowledgements}

%\tableofcontents

\bibliographystyle{aa}
\bibliography{34514}

\end{document}